\documentclass[structabstract]{aa}  

\usepackage{graphicx}
\usepackage{natbib}
\usepackage{verbatim}
\usepackage{color}
\usepackage{multirow}

\newcommand{\tenpow}[2]{#1$\times$10$^{#2}$}
\newcommand{\valerr}[2]{#1$\pm$#2}
\def\src{NGC~7538~IRS~1}

\def\hd{H\,{\tiny II}}

\def\htcop{H$^{13}$CO$^{+}$}

\def\cdo{C$^{17}$O}
\def\so{SO}
\def\cts{C$^{34}$S}

\def\cmt{cm$^{-3}$}
\def\kms{km~s$^{-1}$}
\def\jybeam{Jy~beam$^{-1}$}

\def\asec{$''\!\!\!.\,$}

\def\deg{$^\circ$}
\def\msun{$M_{\odot}$}
\def\lsun{L$_{\odot}$}

\def\lsim{\mathrel{\rlap{\lower4pt\hbox{$\sim$}}\raise1pt\hbox{$<$}}} 
\def\gsim{\mathrel{\rlap{\lower4pt\hbox{$\sim$}}\raise1pt\hbox{$>$}}} 

\begin{document}
   \title{Shaping a high-mass star-forming cluster through stellar feedback\thanks{Based on observations carried out with the SMA telescope. 
The SMA is a joint project between the Smithsonian Astrophysical 
Observatory and the Academia Sinica Institute of Astronomy and Astrophysics,
 and is funded by the Smithsonian Institution and the Academia Sinica (\texttt{http://sma1.sma.hawaii.edu/}).} }

   \subtitle{The case of the NGC~7538~IRS~1--3 complex}

   \author{P.\ Frau \inst{1,2}\thanks{Part of this project was done
   under the affiliation of Institut de Ci\`encies de l'Espai, CSIC--IEEC, and 
   as a visitor at the Harvard-Smithsonian Center for Astrophysics.}
	\and
	J.~M.\ Girart \inst{3}
	\and
	Q.\ Zhang \inst{4}
	\and
	 R.\ Rao \inst{5}
	   }

\institute{
Instituto de Ciencia de Materiales de Madrid (CSIC), Sor Juana In\'es de la Cruz 3,
E-28049 Madrid, Spain\\
\email{pfrau@icmm.csic.es}
\and
Observatorio Astron\'omico Nacional, Alfonso XII 3, E-28014 Madrid, Spain\\
\email{p.frau@oan.es}
\and
Institut de Ci\`encies de l'Espai, CSIC-IEEC, Campus UAB, 
Facultat de Ci\`encies, Torre C5p~2, E-08193 Bellaterra, Catalonia, Spain\\
\email{girart@ice.cat}
\and
Harvard-Smithsonian Center for Astrophysics, 60 Garden Street, Cambridge MA~02138, USA\\
\email{qzhang@cfa.harvard.edu}
\and
Institute of Astronomy and Astrophysics, Academia Sinica, 645 N. Aohoku Place, Hilo HI~96720, USA\\
\email{rrao@sma.hawaii.edu}
}

   \date{Received  \ldots\ / Accepted \ldots\ / Published \ldots\ }

  \abstract
   {NGC~7538~IRS~1--3 is a high-mass star-forming cluster with several detected dust cores, infrared sources, (ultra)compact \hd\ regions, molecular outflows, and masers. In such a complex environment, important interactions and feedback among the embedded objects are expected to play a major role in the evolution of the region.}
   {We study the dust, kinematic, and polarimetric properties of the NGC~7538~IRS~1--3 region to investigate the role of the different forces interplaying in the formation and evolution of high-mass star-forming clusters.}
   {We perform SMA high angular resolution observations at 880~$\mu$m with the compact configuration. We develop the \texttt{RATPACKS} code to generate synthetic velocity cubes from models of choice to be compared to the observational data. We develop the ``mass balance'' analysis to quantify the stability against gravitational collapse accounting for all the energetics at core scales.}
   {We detect 14 dust cores from 3.5~\msun\ to 37~\msun\ arranged in two larger scale structures: a central bar and a filamentary spiral arm. The spiral arm presents large scale velocity gradients in \htcop~4--3 and \cdo~3--2, and magnetic field segments well aligned to the dust main axis. The velocity gradient is well reproduced by a spiral arm expanding at 9~\kms\ with respect to the central core MM1, which is known to power a large precessing outflow. The energy of the outflow is comparable with the spiral arm kinetic energy, which is dominant over gravitational and magnetic energies. In addition, the dynamical ages of the outflow and spiral arm are comparable. At core scales, those embedded in the central bar seem to be unstable against gravitational collapse and prone to form high-mass stars, while those in the spiral arm have lower masses that seem to be supported by non-thermal motions and magnetic fields.}
   {The NGC~7538~IRS~1--3 cluster seems to be dominated by proto-stellar feedback. The dusty spiral arm appears to be formed in a snow-plow fashion due to the outflow from the MM1 core. We speculate that the external pressure from the red-shifted lobe of the outflow could trigger star formation in the spiral arm cores. This scenario would form a small cluster with a few central high-mass stars, surrounded by a number of low-mass stars formed through proto-stellar feedback.}


   \maketitle

\section{Introduction}

NGC~7538 is a well studied optically visible \hd\ region, located at a distance of  2.65~kpc \citep{moscadelli09}, surrounded by very massive star--forming complexes  \citep{Minn75, Werner79, Read80, Rots81, Campbell84, Reid05}. NGC~7538 IRS~1--3  form a cluster of embedded massive protostars with luminosities of $\sim 10^4$~\lsun\  located 2 arcmin (1.5~pc) southeast of the center of the optical \hd\ region  \citep{Wynn74}. IRS~1, 2 and 3 have luminosities of 8, 5 and $0.6\times10^4$~\lsun,  respectively  and all of them have ultra compact (UC) \ion{H}{ii} regions  \citep{Campbell84, Campbell88}.

IRS~1, whose luminosity is equivalent to a O7.5 main sequence star, is surrounded  by an UC \ion{H}{ii} region with a double-lobed structure along the North--South  direction with a size of 0\asec2 ($\simeq 500$~AU) \citep{Campbell84a, Gaume95}. At these scales, the radio emission is dominated by a collimated, ionized wind that exhibits time variability \citep{Franco04, Sandell09}.  This is supported by the extremely  broad line widths of radio recombination lines \citep{Gaume95,Keto08}.  A bipolar molecular outflow extending in the NW--SE direction is detected in CO \citep[][hereafter QZM11]{Kayema89, davis98, qiu11}.  The outflow has created a cavity that is well  detected in the near and mid-IR at scales of a few arcseconds north of the protostar  \citep{Buizer05, Kraus06}. Despite of the existence of the UC \ion{H}{ii} region,  the protostar is probably still actively accreting gas with a high infall rate  (mm inverse P--Cygni profiles yield infall rates of $\sim 10^{-3}$~\msun~yr$^{-1}$) and it probably has a massive circumstellar disk \citep[QZM11; ][]{Klaassen09, Sandell09, Beuther12}. This source is associated with a rich variety of masers, most of them  arising from the interaction zone of the dense molecular gas with the ionized gas  and from the outflow \citep{Rots81,Dickel82, Johnston89, Schilke91, Minier98,  Hoffman03, Hutawarakorn03,Pestalozzi06, Galvan10, Surcis11,  Hoffman11}.

IRS~2 and IRS~3 are not associated with a dusty envelope and are probably in a more evolved phase. IRS~2 is  associated with an O5 star and it is probably the most evolved source. \citet{bloomer98} propose that its stellar winds are shocking the surrounding material, generating a ``stellar wind bow shock'' visible as a shell in \hd\ and Fe$_{\rm II}$. IRS~3 is associated with an O6--O9 star that might power one or more CO outflows \citep[QZM11;][]{ojha04}.

The dense molecular gas around IRS~1--3 appears to have a filamentary morphology, with an arc--like shape southeast of IRS~1 \citep[QZM11; ][]{Pratap90}. \citet{kawabe92} interpret this dense molecular gas filament as part of an expanding  ring--like structure with a radius of 0.25~pc and a mass of 230~\msun, created or  piled up by the strong protostellar winds. QZM11 presented a study of the NGC~7538~IRS~1--3 region using the Submillimeter Array (SMA) in the 1.3~mm wave band. They found star--forming cores deeply embedded within the filamentary, dense molecular gas cloud of IRS~1--3 and multiple molecular outflows.

Previous submm polarization observations by \citet{Momose01} were done with single--dish bolometers toward IRS~1--3 and IRS~11 (a younger source, located $1\farcm5$ south of IRS~1).  The two sources show striking differences in the polarization properties.  Thus, while IRS 11 exhibits an extremely well-ordered magnetic field and a high degrees of polarization, in IRS~1 the field appears locally  disturbed, and the degrees of polarization are much lower than those of IRS~11. They interpreted this as an evolutionary effect (more ordered fields are observed in younger sources), which has been also observed in other high mass star forming regions \citep{girart09, girart13,Tang09a, Tang09b}.

Polarimetric observations allow us to study the magnetic fields at the relevant scales (100--$10^4$~AU) where the star formation  takes place \citep{Girart99, girart06, Rao09, Rao14, Hull13,liu13,qiu13}.  Recently, \citet{zhang14} presented a large sample of massive, clustered, star forming clumps and showed that magnetic fields play an important role during the formation of dense cores at spatial scales of 0.01--0.1~pc. In this paper, we present SMA observations carried out at 345 GHz toward the massive cluster \src. Section~\ref{sec-obs} briefly describes observations and data reduction. Section~\ref{sec-results} presents  the results of the observations. Sections~\ref{sec-analysis} and \ref{sec-disc} contain  the analysis and discussion. Finally, in Section~\ref{sec-concl} we draw our main  conclusions.

   \begin{figure*}
   \centering
   \includegraphics[width=.7\textwidth,angle=0]{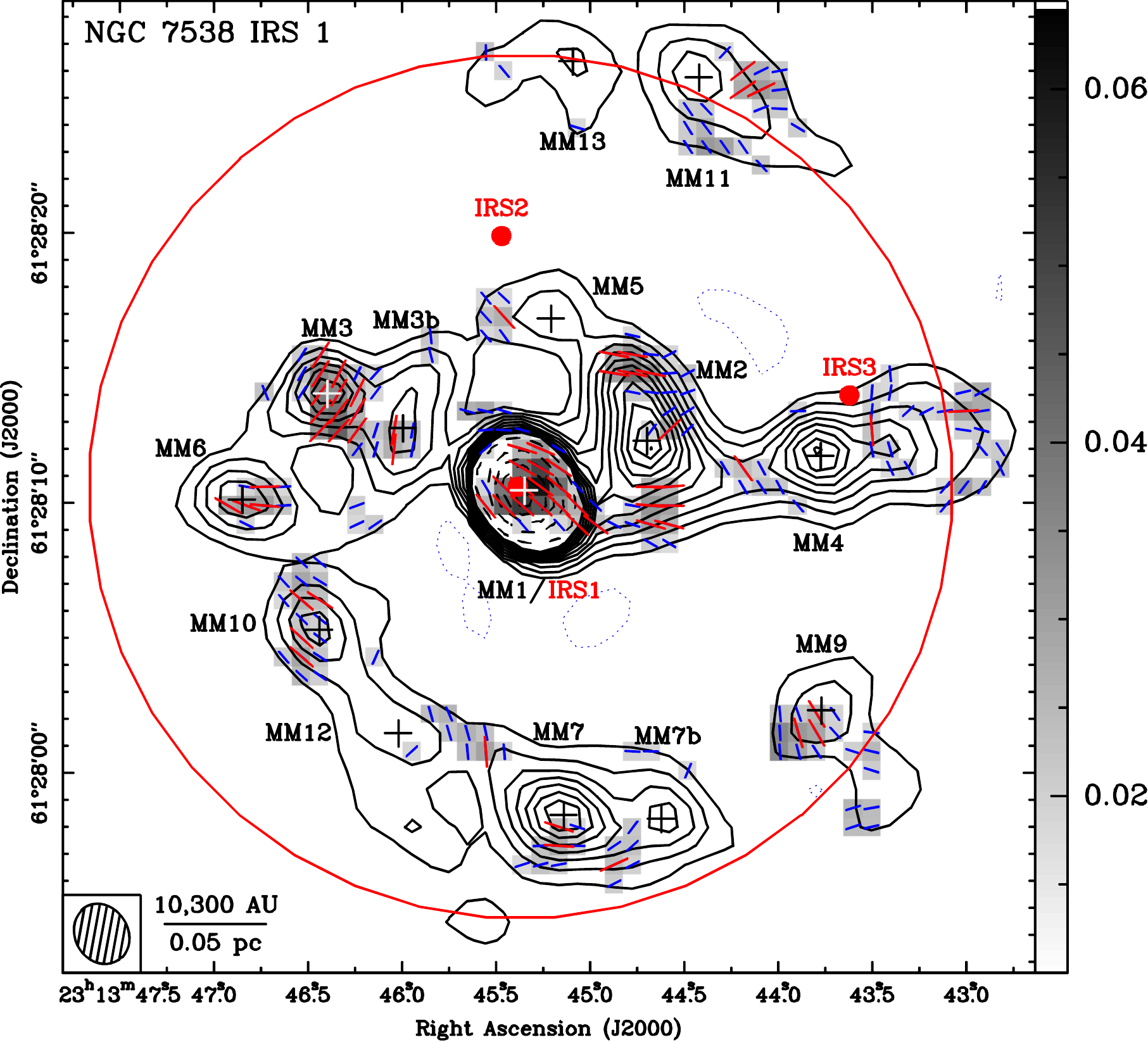}
   
   \caption{Contour map of the SMA dust continuum emission map towards \src\ at 878~$\mu$m overlapped with the grey image of the polarized intensity.  The blue and red segments show the magnetic field direction obtained from the polarization data with cutoff at rms of 2 and 3--$\sigma$, respectively (see Section~\ref{sec-pol}). Solid contours go from 3--$\sigma$ to 33--$\sigma$ in steps of 3--$\sigma$, where  $\sigma=0.017$~Jy~beam$^{-1}$ is the rms noise of dust continuum. The dashed  contours around IRS~1 (MM1) go from 50--$\sigma$ to 300--$\sigma$ in steps of  50--$\sigma$. The scale of the polarized intensity image is shown in the right--hand  side of the figure (the units are Jy~beam$^{-1}$). The black and white crosses show the peak positions of the dust continuum sources (Table~\ref{tab-cont}). The name of the dust continuum sources are also shown. Red circles mark the positions of IRS~1--3. The solid  circle shows the FWHM of the SMA primary beam at the observed frequency.  The physical scale of the map and synthesized beam are shown in the bottom left corner of the panel.}
   \label{fig-pol}
    \end{figure*}

\begin{table*}

\caption{
878~$\mu$m continuum emission parameters.
}
\begin{tabular}{cccccccccc}
\hline
&
\multicolumn{1}{c}{$\alpha$(J2000)} &
\multicolumn{1}{c}{$\delta$(J2000)} &
\multicolumn{1}{c}{$I_\nu^{\textrm{\tiny{Peak}}}$} &
\multicolumn{1}{c}{$S_{\nu}$} &
\multicolumn{1}{c}{$\diameter_{\rm Peak/2}$ $^b$} &
\multicolumn{1}{c}{$T_{\rm dust}$ $^a$} &
\multicolumn{1}{c}{$N_{\rm H_2}$ $^c$} &
\multicolumn{1}{c}{$n_{\rm H_2}$ $^c$} &
\multicolumn{1}{c}{Mass $^c$} 
 \\
\multicolumn{1}{c}{Source} &
\multicolumn{1}{c}{h:m:s} &
\multicolumn{1}{c}{$^\circ$:$'$:$''$} &
\multicolumn{1}{c}{(\jybeam)} &
\multicolumn{1}{c}{(Jy)} &
\multicolumn{1}{c}{(10$^{3}$AU)} &
\multicolumn{1}{c}{(K)} &
\multicolumn{1}{c}{(10$^{23}$cm$^{-2}$)} &
\multicolumn{1}{c}{(10$^{6}$cm$^{-3}$)} &
\multicolumn{1}{c}{($M_{\odot}$)} 
\\
\hline
MM1	&23:13:45.349	&61:28:10.47    &5.24	&\valerr{6.07}{0.05}	&4.93 	&245 	&14 	&23    	&17 \\
MM2	&23:13:44.698	&61:28:12.30    &0.563	&\valerr{1.87}{0.08}	&9.38 	&40  	&8.4   	&7.5   	&37 \\
MM3	&23:13:46.397	&61:28:14.05    &0.358	&\valerr{0.52}{0.05}	&6.04  	&40  	&5.6   	&7.8   	&10 \\
MM3b	&23:13:45.976	&61:28:13.52    &0.214	&\valerr{0.57}{0.06}	&9.54  	&40  	&2.5   	&2.2   	&11 \\
MM4	&23:13:43.774	&61:28:11.74    &0.419	&\valerr{1.30}{0.06}	&8.36  	&40  	&7.4   	&7.4   	&26 \\
MM5	&23:13:45.207	&61:28:16.83    &0.149	&\valerr{0.18}{0.04}	&6.28  	&40  	&1.8   	&2.4   	&3.5 \\
MM6	&23:13:46.849	&61:28:10.12    &0.251	&\valerr{0.39}{0.05}	&6.75  	&43  	&3.1   	&3.8   	&7.0 \\
MM7	&23:13:45.140	&61:27:58.44    &0.404	&\valerr{0.97}{0.07}	&7.60  	&58  	&4.3   	&4.7   	&12 \\
MM7b	&23:13:44.622	&61:27:58.30    &0.224	&\valerr{0.37}{0.05}	&6.75  	&58  	&2.1   	&2.6   	&4.7 \\
MM9	&23:13:43.770	&61:28:02.32    &0.153	&\valerr{0.31}{0.05}	&8.36  	&54  	&1.2   	&1.2   	&4.2 \\ 
MM10	&23:13:46.439	&61:28:05.30    &0.229	&\valerr{0.36}{0.05}	&6.52  	&43  	&3.1   	&3.9   	&6.5 \\ 
MM11	&23:13:44.421	&61:28:25.76    &0.194	&\valerr{0.50}{0.05}	&8.17  	&40  	&2.9   	&3.0   	&9.6 \\	
MM12	&23:13:46.019	&61:28:01.47    &0.137	&\valerr{0.32}{0.04}	&9.54  	&43  	&1.3   	&1.1   	&5.6 \\ 
MM13	&23:13:45.092	&61:28:26.36    &0.105	&\valerr{0.18}{0.04}	&8.71  	&40  	&0.94  	&0.90   	&3.5 \\	

\hline
\end{tabular}
\smallskip\\
$^a$ For the sources detected by QZM11 we use the temperature they use. For the rest of the sources we adopt 40~K.\\
$^b$ Diameter of the circle with area equal to the source area satisfying $I_\nu$$>$$I_\nu^{\rm Peak}/2$\\
$^c$ Assuming $\kappa_{\rm 341.4~GHz}$$=$0.015~cm$^2$~g$^{-1}$ \citep{ossenkopf94}. See Appendix~1 in Frau et al. (2010) for details on the calculation. The uncertainties are approximately 50\% of the computed values. \\
\label{tab-cont}
\end{table*}

\section{Observations and data reduction\label{sec-obs}}

The SMA observations were undertaken on 2005 July 7 with seven  antennas in the compact configuration. The weather was good during the observations,  with system temperatures in the range of 200--300~K. The phase center used for \src\ was $\alpha$(J2000.0)$= 23^{\rm h}13^{\rm m}43\fs359$, $\delta$(J2000.0)$= 61\degr 28' 10\farcs60$ \citep{davis98}.  A single receiver was used and tuned to a Local Oscillator frequency of 341.6 GHz (878~$\mu$m), with a total bandwidth of 2~GHz per  sideband, covering a frequency range of 335.51--337.49~GHz in the lower sideband and  about 345.51--347.89~GHz in the upper sideband.  At these frequencies, the full width at  half maximum (FWHM) of the antennae's primary beam is $32\farcs4$. The 2$\times$2~GHz  correlator was configured to sample the aforementioned frequency ranges with a uniform  spectral resolution of 0.81 MHz ($\simeq0.7$~\kms). The standard reduction of the data  was done by using the IDL MIR package and selecting the QSO 3C454.3 for the bandpass  calibration and BLLAC for the gain calibration. The polarization calibration was performed by observing  3C454.3 over a large parallactic angle range ($-70\degr$ to $+70\degr$). This allowed us to correct for the instrumental polarization using the MIRIAD software package at an accuracy of 0.1\% \citep{Marrone08}.  We used the MIRIAD task UVLIN to separate the  continuum and the line data in the $u,v$ domain. 
The line-free channels were selected by inspecting MM1 because it is the core with the most molecular lines. These channels were used for the final maps, minimizing the possible contamination from molecular line emission. Self--calibration on \src\ was performed using the Stokes $I$ continuum data for each baseline independently. The derived gain solutions were applied  to the molecular line data.  

The final maps were obtained using Natural weighting, which yielded a synthesized beam  of $2\farcs33\times2\farcs01$ with a position angle of $34\degr$. The continuum map sensitivity is $\sigma_{I}$=0.017~\jybeam\ for Stokes~$I$ and $\sigma_{\rm pol}$=0.010~\jybeam\ for Stokes~$Q$ and $U$. For the  molecular line data, we made channel maps with a spectral resolution of  1.4~\kms\ that resulted in a sensitivity of $\sigma$=0.25~\jybeam\ per channel.

\begin{figure*}
\centering
\includegraphics[width=.9\textwidth]{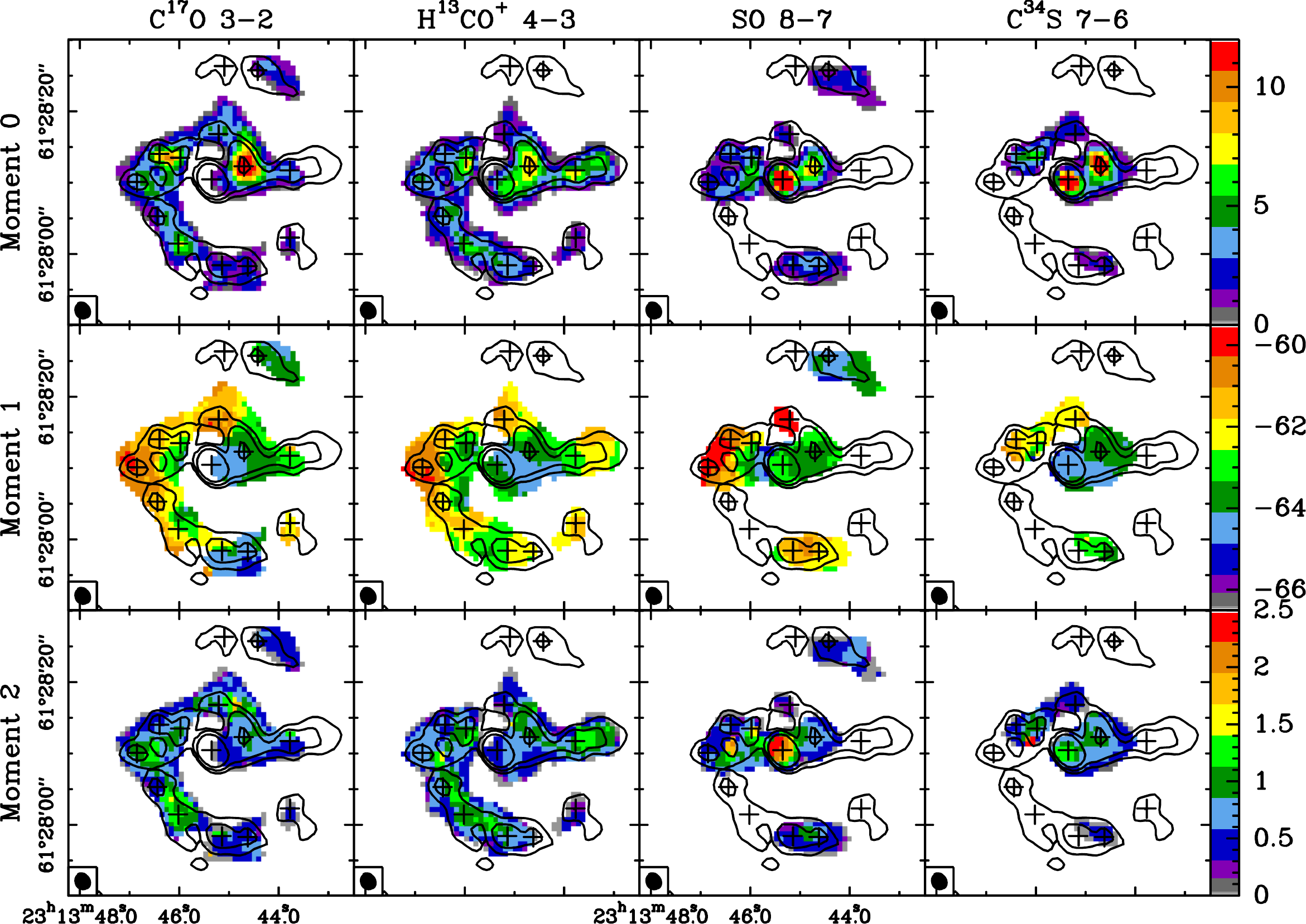}

 \caption{Moment maps of molecular transitions with extended emission. Rows:  moments 0, 1, and 2, in descending order, labeled on the left-hand side of the figure. Columns: from left to right,  \cdo~3--2, \htcop~4--3, \so~8--7, and \cts~7--6, respectively, labeled on the top of the figure. Color map: moment maps. Common scale is shown on the right-hand side of the figure. Contours: 3$\sigma$, 9$\sigma$, and 27$\sigma$ continuum emission levels, where $\sigma$=0.017~Jy~beam$^{-1}$.}

\label{fig-momnt}%
\end{figure*}

\begin{figure*}
\centering
\includegraphics[angle=-90,width=.9\textwidth]{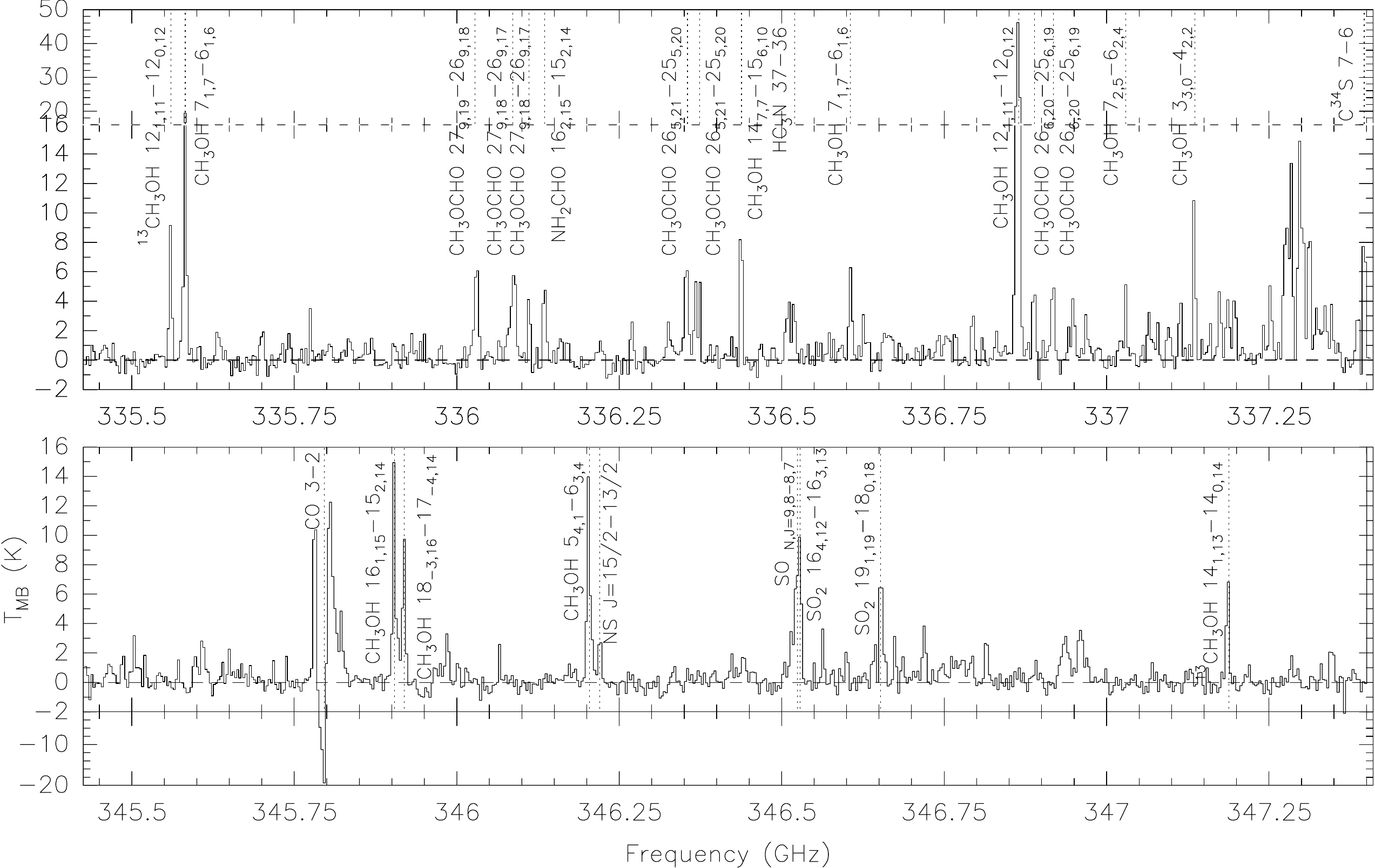}

\caption[MM1 spectrum USB 1 of 2]{Spectrum towards the MM1 peak. Top panel: lower side band. Bottom panel: upper side band. A common temperature range from -2 to 14~K is displayed for better visualization. Additional temperature ranges are shown for the upper (-2 to -20~K) and lower (14 to 42~K) sidebands to display the entire spectrum. A dashed lines marks the 0~K level. For both sidebands, the frequencies of the main molecular transitions are noted by a dotted vertical line and the transition is specified.}

\label{fig-spec}
\end{figure*}

\section{Results\label{sec-results}}

\subsection{Continuum emission}

Figure~\ref{fig-pol} presents the 878~$\mu$m continuum SMA map of \src. A total of fourteen  sources are detected, including IRS~1 (MM1 in this figure). MM1 is the brightest source  at 878~$\mu$m, with an intensity of 5.24~\jybeam. The rest of the continuum sources have  intensities in the 0.11--0.56~\jybeam\ range. Our mass sensitivity at a 3$\sigma$ level for a typical $T_{\rm dust}=40$~K is $\sim$0.9~\msun~beam$^{-1}$, twice as better as that of QMZ11 (our second contour in Fig.~\ref{fig-pol} roughly compares to their first contour in Fig.~1). Hence, only nine sources were previously detected by QMZ11. MM8, detected by QZM11, is not detected  at 878~$\mu$m because it falls outside the FWHM of the  primary beam at this wavelength. The 878~$\mu$m position of MM9 is offset by $4\farcs5$ with respect to the value given at  1.3~mm by QZM11. This is probably due to the relative weakness of this source at the two wavelengths and that it is close to the primary beam edge. MM3 and MM7 show a secondary peak at 878~$\mu$m, MM3b and MM7b, which were not detected previously.  Both new sources are located westward of the main component and have peak fluxes of  $\sim$55\% that of the main component. In addition, four new sources are detected at 878~$\mu$m, named MM10--MM13 following the convention of QZM11.  All the dust continuum sources except MM11 and MM13 appear to be located  in a diffuse arc--like filamentary structure, resembling a ``spiral arm'' which encloses IRS~1 (hereafter we refer to this filament as ``spiral--arm'').  This ``spiral arm'' was first reported by \citet{kawabe92} from CS~2--1 observations.  At the center of the complex, MM1 and MM2 sources seem to be closer to each other and embedded in a dense environment. The other sources, located mainly east and south of IRS~1, are embedded in more diffuse medium and form a C-shaped structure that seems to arise from MM1-MM2.

The measured and derived physical parameters of the sources are listed in Table~\ref{tab-cont}. Temperatures are assumed to be those used by QZM11:  245~K for MM1 and 40--58~K for the rest of the sample. For the new sources (MM10--13),  we adopted a temperature of 40~K (these sources are weaker and relatively far from  IRS~1).  The diameter of the sources (FWHM of the dust emission) are relatively  homogeneous with an average value of 3\asec0$\pm$0\asec5 ($\simeq 8 \times 10^3$~AU),  slightly larger than the beam, and thus are poorly resolved. The densest object is MM1 (\tenpow{2.8}{7}~\cmt) but the most massive one is MM2 ($\simeq 37$~\msun). The total mass is $\simeq$160~\msun, where 50~\% is contained in the MM1-MM2-MM4 region ($\simeq 80$~\msun) and 40~\% in the C-shaped filament SE to MM1 ($\simeq 60$~\msun). In addition, the central sources appear to be denser than those in the C-shaped filament, with the exception of MM3.

\subsection{Linearly polarized continuum emission\label{sec-pol}}

The polarized emission is broadly detected along the dust filamentary structure, with a polarized intensity between $\simeq 0.017$ and 0.059~\jybeam. Two sets of polarization segments were computed. Firstly, the high--significance set (red segments in Fig.~\ref{fig-pol}) is computed using 3.0--$\sigma_{\rm pol}$ cutoff in the Stokes $Q$ and $U$ maps and 6--$\sigma_{\rm I}$ cutoff in the Stokes $I$ maps. Secondly, the low--significance set (blue segments in Fig.~\ref{fig-pol}) is computed using 2--$\sigma_{\rm pol}$ and 3--$\sigma_{\rm I}$ cutoffs, respectively. The agreement in the magnetic field direction is remarkable between both sets, thus the low cutoff values deliver realistic information (Section~\ref{ssec-houde09}). The overall morphology of the magnetic field segment directions\footnote{The so called magnetic field segments, represent the angle of the line-of-sight (LOS) integrated linearly polarized emission flipped by 90\deg, which is assumed to roughly trace the magnetic field direction.}  is not uniform across the region, unlike the ordered directions detected in other regions \citep{girart06,girart09}, and seem to roughly follow the arc--like filament direction as traced by the continuum emission.

\begin{figure*}
\centering
\includegraphics[angle=0,width=.24\textwidth]{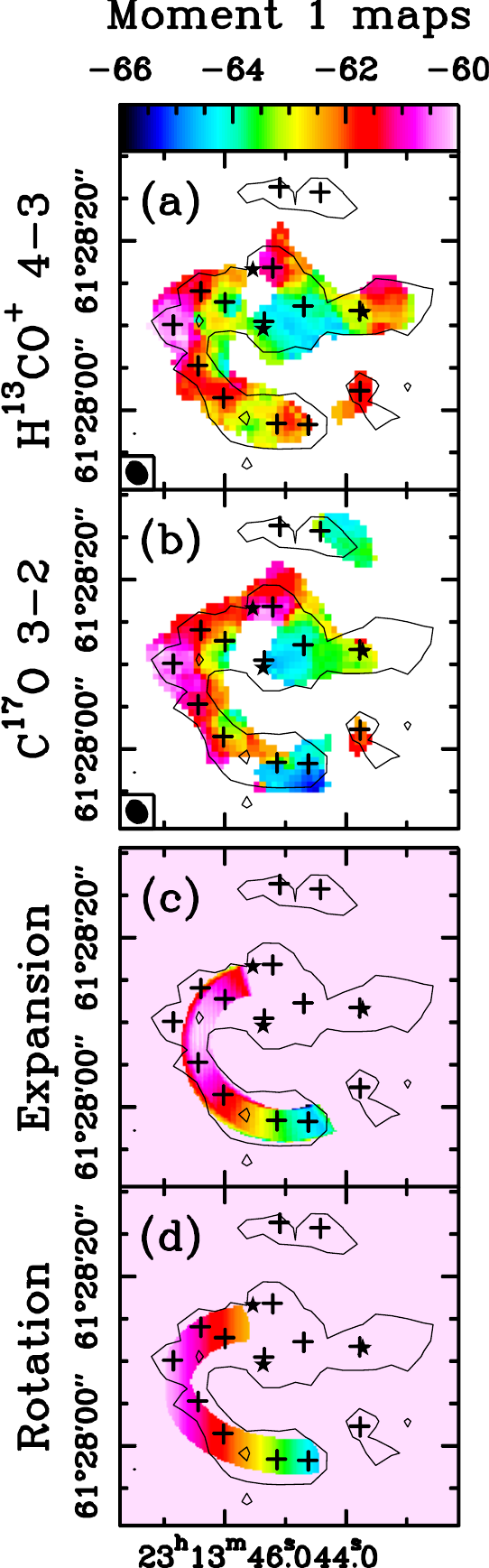} 
\hfill
\includegraphics[angle=0,width=.74\textwidth]{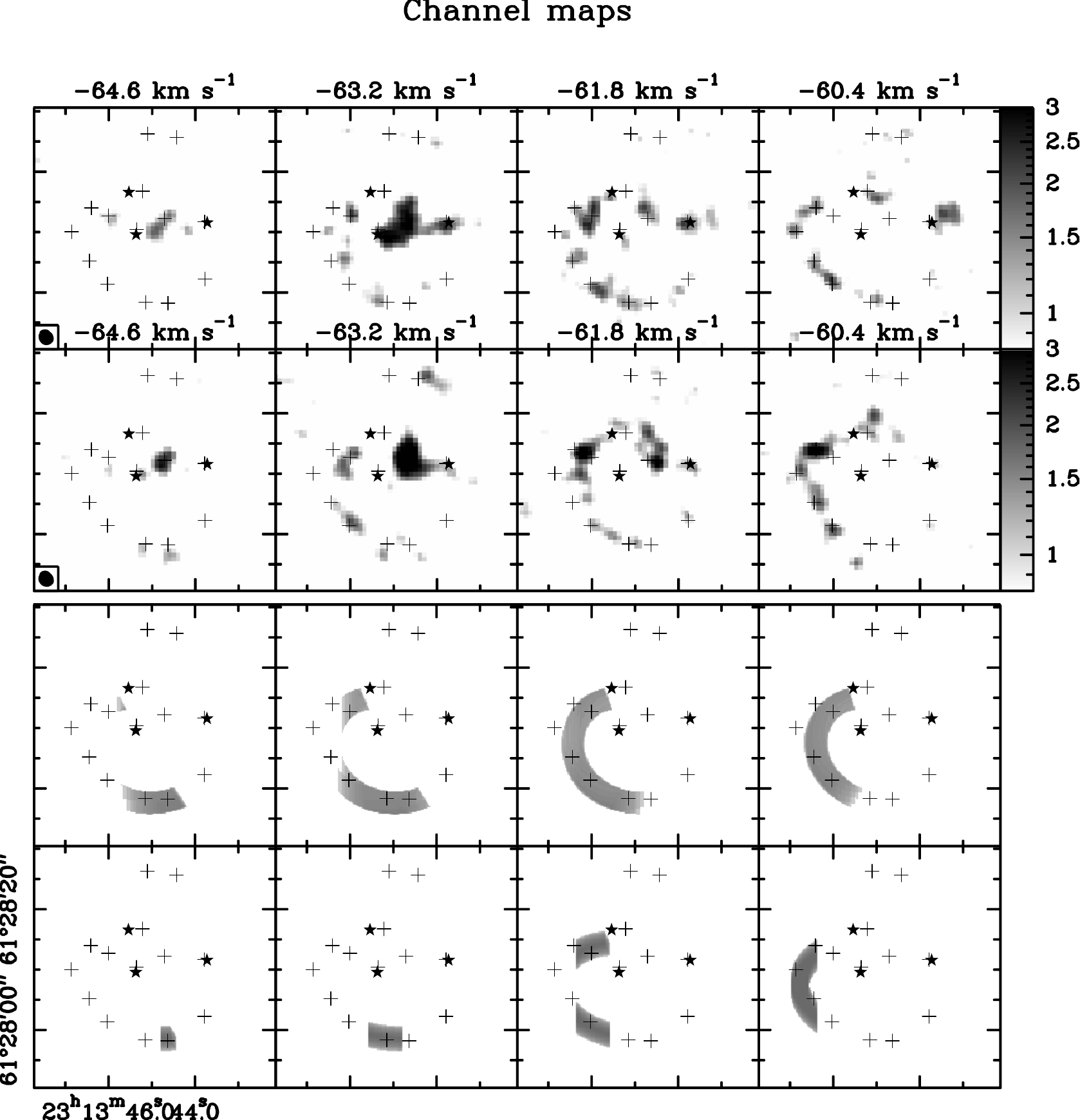} 

\caption{Velocity structure of \src.  {\it Contours}: observed dust continuum map, contours are 3$\sigma$ to 21$\sigma$ in steps of 6$\sigma$, where $\sigma$=0.025~\jybeam. {\it First column}: order~1 moment maps, i.e. velocity structure. The common scale is shown at the top of the column. {\it Second to fifth columns}: channel maps with the velocity labeled at the top. {\it Rows a and b}: observed \htcop~4--3 (a) and \cdo~3--2 (b) maps. The gray-scale for the channel maps is shown at the right-hand side of the figure.  {\it Rows c and d}: synthetic maps generated with \texttt{RATPACKS} (Section~\ref{ssec-toy-model}) for a logarithmic spiral with radial expanding motions (c), and with rotational motions (d).}
\label{fig-toymodel}
\end{figure*}

\subsection{Molecular line emission}

In this section we present the molecular line data towards \src\ at 878~$\mu$m that can be summarized in two different behaviors: ({\it i}) four molecular transitions that have extended emission arising from the diffuse material, and ({\it ii}) many spatially unresolved hot-core lines only present towards the chemically differentiated MM1 source.

The four molecular transitions that exhibit extended emission are shown in Figure~\ref{fig-momnt}, which presents the zero, first and second order moment maps (integrated intensity, velocity and velocity dispersion maps, respectively) of the emission. The different transitions are ordered by increasing critical density\footnote{$n_{\rm crit}=A/\gamma$ with the Einstein coefficient $A$ and the collisional rate $\gamma$ were taken from LAMBDA \citep{schoier05} for a temperature of 40--50~K.}: $\sim$10$^{5}$~\cmt\ for \cdo~3--2, a few 10$^{6}$~\cmt\ for \htcop~4--3 and \so~8--7, and $\sim$10$^{7}$~\cmt\ for \cts~7--6. These moment maps cover the  velocity range between $-66$ and $-60$~\kms, where the emission is detected. The two most extended molecular transitions, \cdo~3--2 and \htcop~4--3 (first and second column, respectively) trace the dust emission with high fidelity. For both transitions, the first order moment maps show an almost equivalent velocity pattern, which strongly suggest that the complex velocity structure is real. MM1-MM2 sources have a velocity around $-64.5$~\kms, while the rest of the sources seem to be closer to $-61$~\kms. These features are not so well observed in the \cdo\ because it does not trace very well MM1 and MM4. The ``spiral arm'' possibly starting in the MM1-MM2 region and ending up in MM9, is very well traced by the \cdo\ and \htcop\ emission. The projected velocity pattern shows an increase in velocity from MM1 to MM6, and then a slight decrease along the filament down to MM7. An interesting feature of this filament is that appears to be a marginal velocity gradient across the filament, with the inner edge of the arc--like structure blueshifted with respect to the outer edge. 

The \so~8--7 and \cts~7--6 emission (third and fourth column, respectively, of Fig.\ref{fig-momnt}) appears to be less extended than that of the \cdo\ and \htcop\ lines, tracing only partially the dusty arc--like filament. This can be due to their higher critical density, i.e. they trace the densest parts of the filament. Both are present around MM1 (including MM2, MM3, MM3b and MM5) and MM7. The SO line is also detected toward MM6 and MM11. MM1 is particularly bright in these two molecular transition maps and shows a large velocity dispersion of 2--3~\kms, in contrast with the rest of the filament that hardly achieves a dispersion of 1.5~\kms. The velocity pattern of the SO and \cts\ is compatible with that of the more extended, \cdo\ and \htcop\ lines.

The set of molecules detected only towards MM1 continuum emission peak are all spatially unresolved. Figure~\ref{fig-spec} presents the spectrum towards the peak of MM1 with most of the lines labeled. This spectra is typical of an evolved hot-core (see \S~\ref{sec-disc}):  it contains a few transitions of \so, SO$_{2}$, $^{34}$SO$_{2}$, NS, HC$_{3}$N, H$_{2}$CO, and NH$_{2}$CHO; as well as many transitions of CH$_{3}$OH and CH$_{3}$OCHO. The upper energies of these transitions span an order of magnitude, ranging from $\sim$50~K to $\sim$500~K. Several unidentified lines were detected as well.

\begin{table}
\caption{Velocity parameters of the kinematic models}
\begin{center}
\begin{tabular}{c ccccc}

\hline
&
\multicolumn{4}{c}{Velocity}
\\

Fig.~\ref{fig-toymodel}&
Type & $v(r_{\rm out})$ & $r_{\rm out}$ & $\alpha$ & $\theta_{\rm LOS}^{\rm spir}$
\\
(row)&
& \kms & AU & & deg\\
\hline

(c) &
Expansion & 9 & \tenpow{1.5}{4} & 0 & 70 \\
(d) &
Rotation & 2 & \tenpow{1.5}{4} & 1 & 45 \\

\hline

\end{tabular}
\end{center}

\label{tab-toymodel}
\end{table}

\section{Analysis\label{sec-analysis}}

\subsection{The kinematics of the ``spiral arm'' around \src\ \label{ssec-kinetic}}

The velocity pattern found towards the ``spiral arm'' of \src\ is consistent among the tracers and shows very smooth variations (Fig.~\ref{fig-momnt}). This pattern is present towards the south and east of the UC \hd\ region and includes MM3, MM3b, MM5, MM6, MM7, MM7b, MM10, and MM12. The total amount of  mass in the ``spiral arm'' from dust continuum is $\simeq 60$~\msun.

\subsubsection{Kinematic models\label{ssec-toy-model}}

We used a set of simplified models with different geometries and different velocity structures.  For the geometrical structure, we used two-dimensional Archimedean spirals $r = a + b \theta $ and logarithmic spirals $r = a e^{\theta/b} $, both expressed in polar coordinates. A certain thickness was applied to the spiral to form a tubular three-dimensional structure.

For the kinematics, we took into account radial and rotational motions, both with speed following a potential law of the form $v(r)_{rad/rot} = v_{rad/rot}(r_{\rm out}) \left[r/r_{\rm out}\right]^{\alpha}$. At any point in the space, the radial velocity vector is tangential to the radial direction with respect to the source center, while the rotation velocity vector is simultaneously perpendicular to the rotation axis and the radial direction. Positive $v(r_{\rm out})$ mean expansion and counter-clockwise rotation for the cases of radial and rotational motions, respectively. However, changing the signs of the angle with respect to the LOS and of the velocity would produce the same map and, therefore, there is an uncertainty on the direction of the gas flow. The parameter $\alpha$ can be used in radial motions to accelerate or decelerate the gas as a function of the radius. For rotating motions, $\alpha$ can be used to simulate rigid body rotation, constant speed rotation, and keplerian rotation with $\alpha=1$, $\alpha=0$,  and $\alpha=-1/2$, respectively.

Finally, we developed a simple RAdiative Transfer Package for Adaptable Construction of Kinematical and Structural models (\texttt{RATPACKS}) to generate synthetic velocity cubes using any combination of geometric and kinematic input as a synthetic source. The synthetic source can be rotated around any axis allowing any orientation in the three-dimensional space. The velocities are projected on the plane-of-the-sky (POS) according to the velocity pattern chosen, and to the three-dimensional orientation given to the synthetic source.  Then, a simple radiative transfer routine assuming optically thin emission is used to derive noiseless synthetic channel maps, and order~0, 1, and 2 moment maps.

\subsubsection{Application to \src\label{sssec-toy-model-ngc7538}}

We explore the parameter space in order to reproduce best the \htcop\ and \cdo\ maps of the entire ``spiral arm''. We adopt a depth of the spiral arm equal to the average observed width: $5''$ (64~mpc).  Since we are only interested in the general kinematics, we assume uniform gas density, constant molecular abundance and optically thin emission. In Fig.~\ref{fig-toymodel} we present the models with the best fitted parameters together with the \htcop\ and \cdo\ data for comparison. Logarithmic spirals fit best the data than Archimedean spirals and are the ones presented in Fig.~\ref{fig-toymodel}. For the models presented we used $a=$\tenpow{1.4}{4}~AU and $b=42$~rad. The parameters used for the kinematics are listed in Table~\ref{tab-toymodel}. For both velocity cases we set $v_{\rm LSR}=-63$~\kms\ and $r_{\rm out}=$\tenpow{1.5}{4}~AU ($\sim$5\asec7). $\theta_{\rm LOS}^{\rm spir}$ represents the angle between the plane containing the spiral and the LOS, where 0\deg\ and 90\deg\ mean edge-on and face-on, respectively.

Row $c$ in Fig.~\ref{fig-toymodel} shows a synthetic source with radial expanding motions, while row $d$ shows counter-clockwise rotating motions. The radial expanding motions seem to reproduce best the channel maps in both molecules. This velocity pattern produces channel maps with extended emission. The best fit is achieved using $\alpha=0$ with a constant radial velocity of 9~\kms\ and $\theta_{\rm LOS}^{\rm spir}$=70\deg, close to face-on. The rotational pattern produces channel maps with concentrated emission, not seen in the data. The best fit is achieved with a rigid-body rotation of 2~\kms\ at $r_{\rm out}$ and $\theta_{\rm LOS}^{\rm spir}$=45\deg. The values of the velocities depend strongly on the orientation angles of the synthetic source and, therefore, velocities should be taken as upper limits.

\subsection{Statistical derivation of the magnetic field strength\label{ssec-houde09}}

\begin{figure*}[ht]
\centering
\includegraphics[angle=270,width=.9\textwidth]{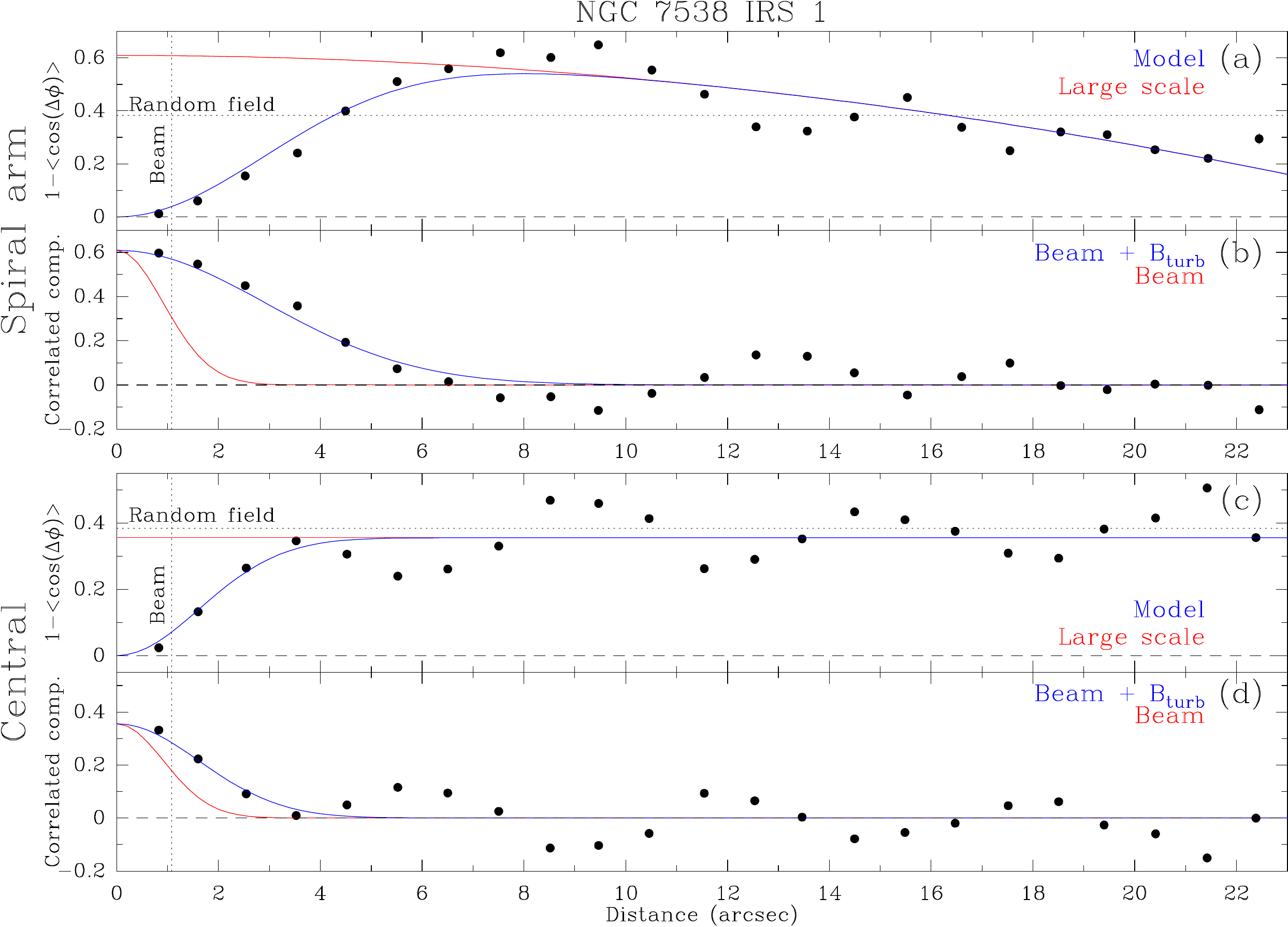}

\caption{Angular dispersion function of the magnetic field segments detected towards the spiral arm (panels a and b) and the central region (panels c and d). {\it Top sub-panels (a and c):} dots represent the data with uncertainty bars, dashed line marks the zero value, dotted vertical line notes the beam size, dotted horizontal line shows the expected value for a randomic magnetic field, red line shows the best fit to the large scale magnetic field (summation in Eq.~\ref{eq-houde09}), and blue line shows the best fit to the data (Eq.~\ref{eq-houde09}). {\it Bottom sub-panels (b and d):} dots represent the correlated component of the best fit to the data, dashed line marks the zero value, dotted vertical line notes the beam size, red line shows the correlation due to the beam, and  blue line shows the correlation due to the beam and the turbulent component of the magnetic field. \label{fig-houde09}}
\end{figure*}

\subsubsection{Formalism}

Based on observational data, a widely used method to estimate the magnetic field strength of the POS component of the large-scale magnetic field is the Chandrasekhar-Fermi \citep[hereafter CF;][]{chandrasekhar53} equation

\begin{equation}
\frac{\delta B}{B_0} \simeq \frac{\sigma_v}{V_{\rm A}},
\label{eq-chand-fermi}
\end{equation}

\noindent where $B_0=\left|\vec{B}_0\right|$ is the large-scale component of the magnetic field, $\delta B$ is the variation about $B_0$, $V_{\rm A}$=$B_0/\sqrt{4\pi\rho}$ is the Alfv\'en speed at density $\rho$, and $\sigma_{\rm v}$ is the velocity dispersion of an appropriate spectral line. Recently, different statistical methods have been developed to avoid some of the CF method caveats \citep{hildebrand09,houde09,houde11,koch10}. These methods rely on the study of the extended polarized emission in observational maps. 

\citet{houde09} assume two statistically independent components of $\vec{B}$, the large-scale magnetic field $\vec{B}_0(\vec{x})$, and the turbulent magnetic field $\vec{B}_{\rm t}(\vec{x})$. Then, they derive the turbulent to large-scale magnetic field strength ratio from the angular dispersion function that accounts for the polarization angle differences as a function of the distance between the measured positions. The analysis is based in an analytical derivation for a turbulent cloud (see their Eq.~4) including the effect of beam and LOS averaging. They further assume a stationary, homogeneous, and isotropic magnetic field strength, an isotropic but negligible turbulent polarized emission, and a magnetic field turbulent correlation length $\delta$ much smaller than the thickness of the cloud $\Delta'$ ($\delta\ll\Delta'$). Applying all these simplifications, the angular dispersion function can be written as

\begin{eqnarray}
1-\langle \cos\left[\Delta\Phi\left(\,l\,\right)\right] \rangle
&\simeq&
\frac{\langle B^2_{\rm t} \rangle}{\langle B^2_0 \rangle}
\,\frac{1}{N}\, 
\left[1-e^{-l^2/2(\delta^2+2W^2)}\right] \nonumber \\
&&+\sum^{\infty}_{j=1}a'_{2j}\,l^{\,2j} ,
\label{eq-houde09}
\end{eqnarray}

\noindent where 

\begin{equation}
N=\left[\frac{(\delta^2+2\,W^2)\Delta'}{\sqrt{2\,\pi}\,\delta^3}\right]
\label{eq-nturb}
\end{equation}

\noindent is the number of independent turbulent cells along the LOS, $W$ is the standard deviation ($\sigma={\rm FWHM}/\sqrt{8\ln{2}}$) of the gaussian beam, and the summation is a Taylor expansion representing the large-scale magnetic field component that does not involve turbulence. The first term in the square brackets contains the integrated turbulent magnetic field contribution, while the exponential term represents the correlation by the combined effect of the beam ($W$) and the turbulent magnetic field ($\delta$).  The intercept of the fit to the data of the non-correlated part at $l=0$, $f_{{\rm NC}}(0)$, together with the assumption of a cloud thickness $\Delta'$, allow us to estimate the turbulent to large-scale magnetic field strength ratio as 

\begin{equation}
\frac{\langle B^2_{\rm t} \rangle}{\langle B^2_0 \rangle}
=
N\,f_{{\rm NC}}(0).
\label{eq-btb0}
\end{equation}

\noindent Finally, identifying $\langle B^2_{\rm t} \rangle \equiv \delta B^2$, one can apply the CF equation (Eq.~\ref{eq-chand-fermi}) to derive the large-scale component of the magnetic field as

\begin{equation}
\langle B^2_0 \rangle^{1/2}=
\sqrt{4\,\pi\,\rho}\,\sigma_v\,
\left[\frac{\langle B^2_{\rm t} \rangle}{\langle B^2_0 \rangle}\right]^{-1/2}.
\label{eq-b0}
\end{equation}

\noindent We note that the magnetic field component labeled as ``turbulent'' describes, more generally, any contribution to the total magnetic field other than the uniform large-scale one. Therefore, when we refer in the next sections to the ``turbulent'' magnetic field we are discussing the non-uniform magnetic field contribution.

\begin{table}
\caption{Derived magnetic field strength~$^a$}

\begin{center}
\begin{tabular}{ l cc}

\hline

&Spiral~$^{b}$ & Central~$^{c}$ \\

\hline
\smallskip

$\delta$  (\arcsec, mpc) & 2.6$\pm$0.3 (33$\pm$4)  & 1.0$\pm$0.6 (13$\pm$8)   \\
$f_{{\rm NC}}(0)$ & 0.61$\pm$0.04 & 0.356$\pm$0.017  \\
$a'_2$ (\arcsec$^{-2}$) &  \tenpow{(-8.5$\pm$1.3)}{-4} & -- \\
 $N$ & 0.9$\pm$0.3  & 6$\pm$8  \\
 $\langle B^2_{\rm t} \rangle / \langle B^2_0 \rangle$ & 1.47$\pm$0.16 & 3.2$\pm$2.8 \\
 $\langle B^2_0 \rangle^{1/2}$~$^{c}$ (mG) & 2.64$\pm$0.14 & 2.3$\pm$1.0 \\

\hline
\end{tabular}
\end{center}

$^a$ Following \citet{houde09} method. We assumed $\Delta'$=5\arcsec$\pm$1\arcsec\ (64$\pm$13~mpc), roughly the width of the filament and the central structure. \\
$^b$ Assuming $\rho$=4$\times$10$^6$~\cmt\ and $\sigma_{v}$=2.3~\kms\ (\htcop~4--3).\\
$^c$ Assuming $\rho$=8$\times$10$^6$~\cmt\ and $\sigma_{v}$=2.2~\kms\ (\htcop~4--3).

\label{tab-houde09}
\end{table}

\subsubsection{Application to \src\label{sssec-houde09-ngc7538}}

The magnetic field segments in this complex region do not follow a defined homogeneous pattern along the observed field (see Fig.~\ref{fig-pol}) such as, e.g., the hour-glass shape reported and modeled in simpler sources \citep{girart06,girart09,frau11,padovani12}. No analytical models are available that can be compared to this complex source. Thus, in order to extract physical information, a statistical approach seems to be the best strategy. The ``spiral arm'' and the central sources seem to have different kinematics and different directional patterns of the segments, probably related to the YSO embedded in the central region (Sec.~\ref{ssec-kinetic} and Fig.~\ref{fig-pol}). Consequently, we analyze the magnetic field for each of the regions independently. Figure~\ref{fig-houde09} shows the angular dispersion function for both structures. Bins are equally spaced by 1\arcsec. Data points represent the mean within the bin, with uncertainties that are smaller than the point size. We used the nonlinear least-squares Marquardt-Levenberg algorithm to fit Eq.~\ref{eq-houde09} to the data. The best fit is shown in Fig.~\ref{fig-houde09} and the parameters are listed in Table~\ref{tab-houde09}.

Panels (a) and (b) show the results for the spiral arm. The uncorrelated large scale component is fitted with a $j=1$ polynomial following Eq.~\ref{eq-houde09}. The correlated component dominates at small distances ($\sim$5\arcsec--\,6\arcsec\ or $\sim$64\,--\,77~mpc at 2.65~kpc).  The turbulent magnetic field effect in correlating the segments is significantly more important than the beam effect. The correlation length is $\delta$=2\asec6$\pm$0\asec3 (33$\pm$4~mpc at 2.65~kpc), almost three times larger than the beam correlation distance. 

Panels (c) and (d) show the results for the central sources. The correlated component is only important in a distance half that of the spiral arm, and is mostly due to the effect of the beam. For larger distances, the data flattens to a value compatible with a random magnetic field \citep[$(1-\cos(\simeq52^\circ)\simeq0.384$:][see also \citealt{girart13}]{poidevin10}. Therefore, the summation in Eq.~\ref{eq-houde09} (large-scale magnetic field) was dropped in our analysis and only the correlated component was used (see Table~\ref{tab-houde09}). The best fit leads to a turbulent magnetic field correlation length of $\delta$=1\asec0$\pm$0\asec6 (13$\pm$8~mpc at 2.65~kpc), roughly equal to $W\sim$0\asec92 (the telescope beam correlation length, Eq.~\ref{eq-houde09}).

To estimate the magnetic field strength, one has to assume either a certain cloud thickness $\Delta'$ or a certain number of turbulent cells $N$ (see Eq.~\ref{eq-nturb}). Since the magnetic field that we are tracing mostly comes from a filamentary structure, an educated guess is to assume that the thickness is that of the filament width, $5''$ (64~mpc). Under this assumption, we found that the spiral arm contains one turbulent cell along the LOS, while the result for the central sources is poorly constrained to 6$\pm$8 cells. In both cases, the local turbulent field is more important than the large scale, ordered field. However, the central sources have a magnetic turbulence two times more important than the spiral arm. The strength of the field is $\sim$2.5~mG, similar for both regions.

\begin{figure}
\centering
\includegraphics[angle=0,width=8.5cm]{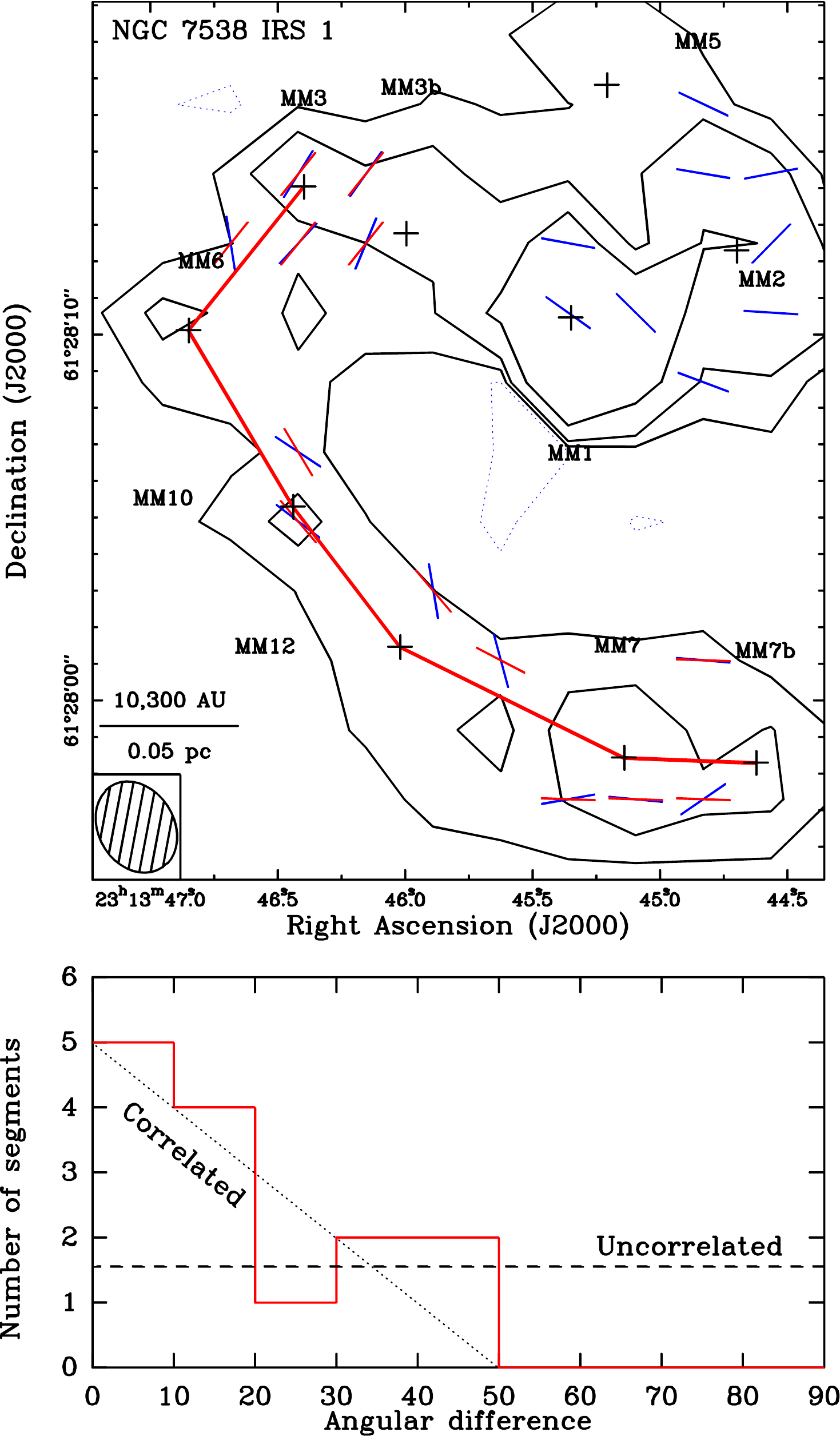}

\caption{Comparison of the dusty filament orientation to the magnetic field segments orientation towards \src. Pixels have been resized to the beam size to ensure statistical independence (see Fig.~\ref{fig-pol} for Nyquist sampling). {\it Top panel}: contours are 3, 9, and 27 times 0.02~Jy~beam$^{-1}$ continuum emission levels. Blue segments are derived as for Fig.\ref{fig-pol}. Red thick line is the axis of the filament (see Section~\ref{ssec-dust-vs-pol}). Red segments show the orientation of the filament corresponding to each pixel with polarization detection. {\it Bottom panel}: histogram of the angle difference between the polarization segment and the filament axis for each pixel. The correlated and uncorrelated distributions used for the $\chi^{2}$ test in Section~\ref{ssec-dust-vs-pol} are shown as dotted and dashed lines, respectively.}

\label{fig-b-dust-ang}
\end{figure}

\subsection{Comparison of dust and magnetic field structures\label{ssec-dust-vs-pol}}

The magnetic field segments seem to roughly follow the direction of the ``spiral arm''. To examine this, we defined the dusty ``spiral arm'' axis as the line connecting the dusty sources (nodes) within the structure (shown in Fig.~\ref{fig-b-dust-ang} as a thick red line). Then, we obtained new maps with pixels of the size of the beam to ensure statistical independence. Each independent pixel with polarized emission was assigned a segment representing the local direction of the filament. This direction was defined as the line connecting the two nearest nodes (red vectors in Fig.~\ref{fig-b-dust-ang}).  Finally, the difference between the magnetic field segment direction and the filament local direction was computed and binned. As shown in the histogram in Fig.~\ref{fig-b-dust-ang}, nine of the fourteen segments ($\sim$64\%) have differences of less than 20\deg\ and none have an angular difference larger than 50\deg. The number of independent measurement is relatively small, therefore, we performed a $\chi^{2}$ test to assess the statistical significance of the results. On the one hand, we compared the data to a flat distribution representing uncorrelated orientations. We found a $<1$\% probability to obtain larger $\chi^{2}$ values with random data, hence the null hypothesis of uncorrelated orientations was rejected. On the other hand, we compared the data to a simple distribution representing correlated orientations. We simplified this distribution to a linearly decreasing function that evolves from total correlation to none in half of the angular range covered (see Fig.~\ref{fig-b-dust-ang}). We found 98\% probability to obtain a larger $\chi^{2}$ value with random data, well above the standard 5\% rejection threshold of the $\chi^{2}$ test, hence the null hypothesis of correlated orientations cannot be rejected. Based on this analysis, we conclude that the orientations of filament and polarization segments are correlated.

\subsection{Energy state of the individual sources: the ``mass balance''\label{ssec-energy-sources}}

We analyze the main causes that are in interplay: gravity, magnetic field, thermal pressure and internal dynamics. On the one hand, gravity has the effect of bringing mass together. On the other hand, the rest of the causes exert the opposite effect either stopping the mass from accreting or dispersing it. We compute a series of meaningful parameters that relate these physical quantities. We also compare the relative strength of these causes in terms of the mass supported against gravitational collapse. 

\subsubsection{Formulae}

We use the viral theorem to check whether the different cores are gravitationally bounded, and  to estimate the maximum mass supported by the thermal and non-thermal motions. These take into account the different internal pressure components. In section~\ref{sec-disc}, we discuss the effect of the external pressure. The virial mass for the thermal component, $M_{\rm T}$, is
\begin{equation}
M_{\rm T}=\frac{k~c_{s}^{2}~R}{G},
\end{equation}
where $c_{s}$, $R$ and $G$  are the sound speed, the core radius and the gravitational  constant, respectively. The parameter $k$ takes into account the specific density distribution of the core. We use $k=1$, which is the value for a density profile $\rho\propto r^{-2}$ \citep{maclaren88}. Similarly, for the non-thermal component the viral mass term, $M_{\rm NT}$, is:
\begin{equation}
M_{\rm NT}=\frac{k~\sigma_{\rm NT}^{2}~R}{G},
\end{equation}
where $\sigma_{\rm NT}$ is the full three-dimensional velocity dispersion of the gas due to non-thermal motions. 

The support of the magnetic field can be included as an additional component in the virial mass. Thus, the mass for a critical mass-to-flux ratio  is given by \citet{nakano78}:
\begin{equation}
M_{\rm mag}=\frac{\pi R^2 B}{\sqrt{4~\pi^{2}~G}}.
\label{eq-m-mag}
\end{equation}
where B is the field strength.

The non-thermal kinetic energy can be compared to thermal kinetic energy by the squared of the turbulent Mach number 
\begin{equation}
\mathcal{M}_{s}^{2}=\left(\frac{\sigma_{\rm NT}}{c_{s}}\right)^{2}.
\label{eq-mach}
\end{equation}
$\mathcal{M}_{s}>1$ means that non-thermal motions are supersonic, and hence, more dynamically important than thermal motions. The thermal energy is compared to magnetic field energy by the plasma $\beta_{\rm T}$
\begin{equation}
\beta_{\rm T}=\frac{P_{\rm therm}}{P_{\rm mag}}=2\left(\frac{c_{s}}{v_{\rm A}}\right)^{2}.
\label{eq-beta}
\end{equation}
where $v_{\rm A}=B_{\rm 3D}/\sqrt{4~\pi~\rho}$ is the  Alfv\'en speed. Similarly, magnetic fields compare to non-thermal motions by 
\begin{equation}
\beta_{\rm NT}=\frac{P_{\rm NT}}{P_{\rm mag}}=2\left(\frac{\sigma_{\rm NT}^{\rm mol}}{v_{\rm A}}\right)^{2}.
\label{eq-beta-turb}
\end{equation}
$\beta_{\rm T}<1$ or $\beta_{\rm NT}<1$ imply that magnetic pressure overcomes thermal or kinetic pressure, respectively.

\begin{figure*}
\centering
\includegraphics[angle=270,width=\textwidth]{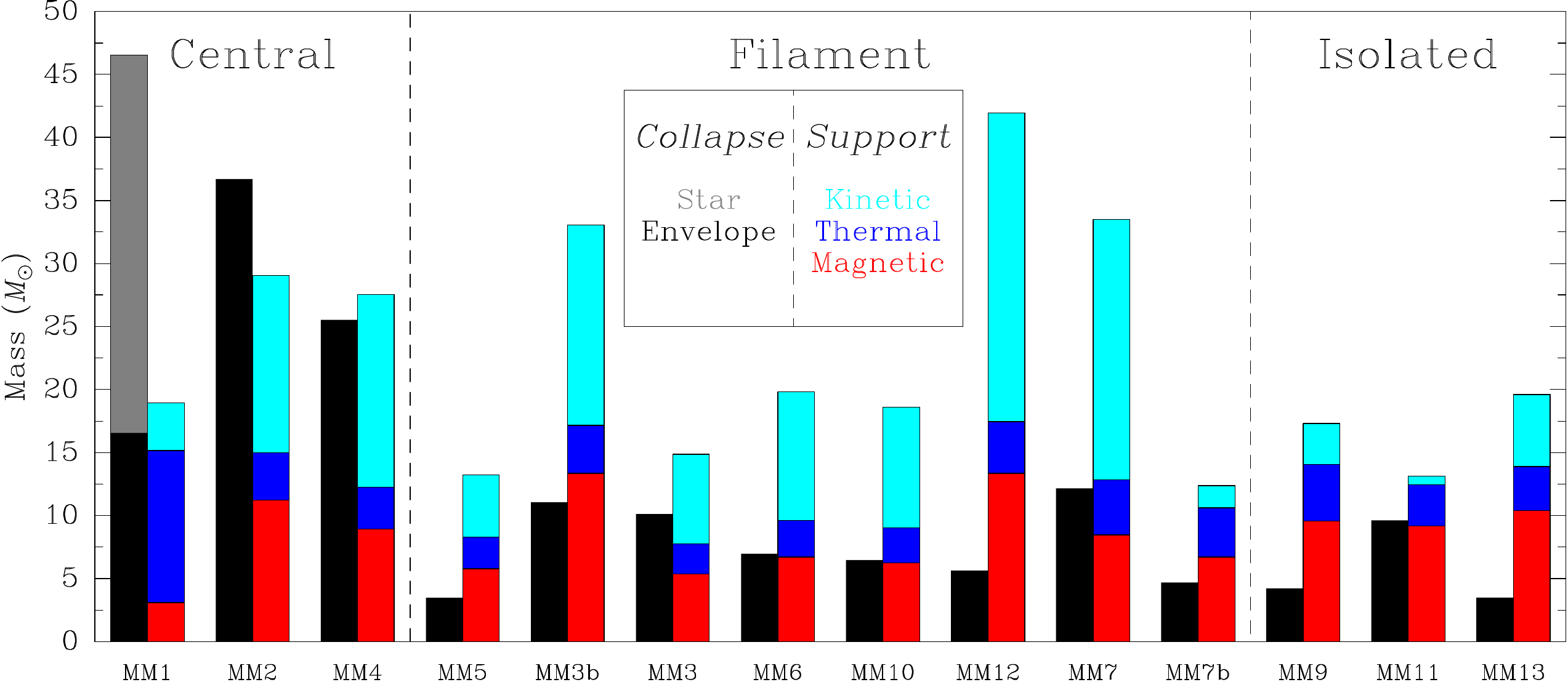} 

\caption{``Mass balance'' analysis. Comparison of the measured mass to the maximum supported mass by different forces. Cores are ordered according to their location in the central massive structure, in the filament, or isolated. {\it Black}: measured mass from continuum maps. {\it Gray}: Mass of the star embedded in the MM1 clump. {\it Red}: mass supported by magnetic fields assuming a uniform value across the source. {\it Light and dark blue}: mass supported by virialized gas motions due to internal dynamics and thermal dispersion, respectively.}

\label{fig-massBalance}
\end{figure*}

\begin{table*}

\caption{
Relative energy indicators and supported masses.
}
\begin{center}
\begin{tabular}{l l ccc ccc cccc}
\hline
\multicolumn{1}{c}{Structure} &
\multicolumn{1}{c}{Source} &
\multicolumn{1}{c}{$c_{s}$} &
\multicolumn{1}{c}{$\sigma_{\rm NT}$} &
\multicolumn{1}{c}{$v_{\rm A}$} &
\multicolumn{1}{c}{$\mathcal{M}_{s}^{2}$} &
\multicolumn{1}{c}{$\beta$} & 
\multicolumn{1}{c}{$\beta_{\rm NT}$} & 
\multicolumn{1}{c}{$M_{\rm T}^a$} & 
\multicolumn{1}{c}{$M_{\rm NT}^b$} & 
\multicolumn{1}{c}{$M_{\rm mag}$} &
\multicolumn{1}{c}{$M_{\rm obs}$}
 \\
\multicolumn{1}{c}{} &
\multicolumn{1}{c}{} &
\multicolumn{1}{c}{\kms} &
\multicolumn{1}{c}{\kms} &
\multicolumn{1}{c}{\kms} &
\multicolumn{1}{c}{} &
\multicolumn{1}{c}{} &
\multicolumn{1}{c}{} &
\multicolumn{1}{c}{\msun} &
\multicolumn{1}{c}{\msun} &
\multicolumn{1}{c}{\msun} &
\multicolumn{1}{c}{\msun} 
\\
\hline

\multirow{3}{*}{Central}
&MM1 
	&1.20&	0.67&	0.96&	0.31&	3.14&	0.98&	12.05&	3.77&	3.10&	16.56 \\ 
&MM2 
	&0.49&	0.94&	1.69&	3.76&	0.16&	0.62&	3.75&	14.07&	11.24&	36.70 \\ 
&MM4 
	&0.49&	1.04&	1.71&	4.58&	0.16&	0.74&	3.34&	15.29&	8.92&	25.54 \\ 
\hline
\multirow{8}{*}{Filament}
&MM5 
	&0.49&	0.68&	3.47&	1.98&	0.04&	0.08&	2.51&	4.96&	5.78&	3.46 \\ 
&MM3b 
	&0.49&	0.99&	3.63&	4.18&	0.04&	0.15&	3.81&	15.91&	13.34&	11.05 \\ 
&MM3 
	&0.49&	0.83&	1.91&	2.95&	0.13&	0.38&	2.41&	7.10&	5.34&	10.11 \\ 
&MM6 
	&0.50&	0.95&	2.73&	3.52&	0.07&	0.24&	2.90&	10.21&	6.70&	6.97 \\ 
&MM10 
	&0.50&	0.93&	2.69&	3.42&	0.07&	0.24&	2.80&	9.58&	6.23&	6.46 \\ 
&MM12 
	&0.50&	1.23&	5.09&	5.98&	0.02&	0.12&	4.10&	24.50&	13.35&	5.64 \\ 
&MM7 
	&0.59&	1.27&	2.46&	4.70&	0.11&	0.53&	4.40&	20.65&	8.46&	12.15 \\ 
&MM7b 
	&0.59&	0.39&	3.33&	0.45&	0.06&	0.03&	3.91&	1.76&	6.70&	4.67 \\ 
\hline
\multirow{3}{*}{Isolated} 
&MM9 
	&0.56&	0.48&	4.50&	0.72&	0.03&	0.02&	4.50&	3.26&	9.56&	4.22 \\ 
&MM11 
	&0.49&	0.23&	2.89&	0.22&	0.06&	0.01&	3.26&	0.70&	9.16&	9.62 \\ 
&MM13 
	&0.49&	0.62&	5.27&	1.64&	0.02&	0.03&	3.48&	5.71&	10.41&	3.50 \\ 

\hline
\end{tabular}
\end{center}
\smallskip
$^a$ To compute the sound speed, $c_{s}=\sqrt{\gamma~k_{\rm B}~T/\mu~m_{\rm H}}$, 
we assume an idealized equation of state with adiabatic index $\gamma = 5/3$ 
\citep{tomida10},  a mean molecular weight of $\simeq2.33$, and the temperature
of the cores estimated by \citet{qiu11}.
\\
$^b$ The 3-D non thermal velocity dispersion is 
$\sigma_{\rm NT}^2 = (\sigma_{obs}({\rm mol})^2 - \sigma_{\rm T}({\rm mol})^{2})$,
where $\sigma_{\rm obs}$ is the observed velocity dispersion 
($\sigma_{\rm obs} = {\rm FWHM} / \sqrt{8~\ln{2}}$) and  $\sigma_{\rm T}$ is 
the thermal line broadening for a molecule of mass $m_{\rm mol}$ 
($\sigma_{\rm T}({\rm mol})=\sqrt{\gamma k_{\rm B}~T/m_{\rm mol}}$). 
\label{tab-energy}
\end{table*}

\subsubsection{Energy ratios in \src\ sources}

The parameters described in the previous section that measure the energy balance among forces are listed in Table~\ref{tab-energy}. Sound speed ranges from 0.49~\kms\ to 0.59~\kms\ except for MM1 that hosts an O7.5 star and is significantly warmer. Non-thermal velocity dispersion $\sigma_{\rm NT}$ range from 0.23~\kms\ to 1.27~kms. The mean value is $0.8\pm0.3$~\kms, $\sim$60\% larger that the typical sound speed at 40~K. As a result, 70\% of the sources show supersonic gas motions $\mathcal{M}_{s}>1$.

We use for each source the magnetic field strength derived in Section~\ref{sssec-houde09-ngc7538}, 2.64~mG and 2.3~mG for the spiral arm and the central sources, respectively. For the isolated cores we used the average of 2.5~mG. This assumption implies that, within each region, the variation of the derived magnetic quantities depends on clump properties: $v_{\rm A}\propto n_{\rm H_{2}}^{-1}$ and $M_{\rm mag}\propto r^{2}$, where $r$ is the clump radius derived as half the diameter from Table~\ref{tab-cont} (see discussion in Section~\ref{ssec-disc-magfields}). The Alfv\'en speed ranges from 0.96~\kms\ to 5.3~\kms, with mean value of $3.0\pm1.1$~\kms. All sources but MM1 have $v_{\rm A}>c_{s}$, and hence, magnetic pressure dominates locally over thermal pressure ($\beta<1$). Non-thermal kinetic energy is comparable to the magnetic energy in four sources: MM1, MM2, MM4, and MM7. For the rest of the sample, magnetic pressure dominates locally over non-thermal pressure.

\subsubsection{``Mass balance'' in \src\ sources}

A similar analysis can be performed in terms of the maximum mass supported by each force, listed in Table~\ref{tab-energy}. The ``mass balance'' accounts for all the available information at once.  MM1 hosts an embedded O7.5 star whose mass is taken into account \citep[30~\msun:][]{pestalozzi04}. In Fig.~\ref{fig-massBalance}, we compare collapse forces versus support forces to derive the individual ``mass balance''. This analysis shows clearly two groups of sources in terms of stability, well correlated with their location in either ({\it i}) the central structure, or ({\it ii}) the spiral arm and isolated.

The sources in the central structure seem have more mass than that which can be supported: MM1 and MM2 have masses larger than the combined maximum supported mass, while in MM4 masses are comparable. In contrast, all sources located either at the filament or in isolation have virial masses larger than measured masses. In general, the main agent against gravity for sources at the filament is the internal dynamics, while for the isolated sources is the magnetic field. See Section~\ref{ssec-collapse} for a discussion on the implications.

\section{Discussion\label{sec-disc}}

\src--3 as a whole presents a complex and rich structure,  velocity field, and magnetic field. In addition, a number of individual cores can be identified. In this section, we first discuss the global properties, and then proceed to analyze the state of the individual cores.

\subsection{Magnetic field properties\label{ssec-disc-magfields}}

\src--3 contains two different regions in terms of magnetic field properties as shown in Section~\ref{ssec-houde09}. These differences are important in two different but related aspects: the relative importance of non-ordered with respect to ordered magnetic fields, and the relative dynamical importance of magnetic fields in the overall picture.

For the central sources, the magnetic field segments are only correlated at distances slightly larger than the correlation distance of the beam. The turbulence of the magnetic field has a mild effect, but the data does not allow us to constrain accurately the contribution. In any case, the transitions to values for the angular dispersion function compatible to a random field happens at a very short distance ($\sim$3\arcsec, $\sim$39~mpc). The number of turbulent cells along the LOS seems to be large, which is a clue that the field is severely distorted. This suggests that non-ordered magnetic fields are more important than ordered magnetic fields. In fact, the ratio of energies $\langle B^2_{\rm t} \rangle / \langle B^2_0 \rangle$ is $\gsim$3. The size of the sources is larger than the correlation distance in all cases, and thus, this region needs to be analyzed source by source. The study of the energetic state of the individual sources suggests that non-thermal motions are dominant. In addition, the magnetic field tends to a random configuration at relatively short scales. These facts suggest that magnetic fields are not important in the overall picture. In summary, the relevance of the magnetic field is small in this region, and the field is mostly non-ordered.

The spiral arm shows different magnetic field properties with respect to the central structure. The correlation length due to the non-ordered magnetic field is accurately determined since it is $\sim$3 times larger than the beam correlation length. The distance between consecutive embedded sources within the filament of 2\arcsec\ (26~mpc) in average, smaller than the magnetic correlation length of 2\asec6 (33~mpc). This implies that the field properties among cores are not independent. Consequently, the filament analysis must take into account the whole complex. According to the statistical analysis, the spiral arm contains only one turbulent cell along the LOS, hence, it has relatively little turbulence. This is in agreement with the well aligned segments observed, and with the sources having non-independent magnetic fields. Finally, the field shows a $\propto l\,^2$ trend at large scales that suggests a smoothly varying field \citep{hildebrand09}. Moreover, the analysis in Section~\ref{ssec-dust-vs-pol} shows that the magnetic field direction is correlated to the dust morphology along the entire 11\arcsec-long filament. All these suggest that the magnetic field has a strong internal coherence within the filament, and that it is somehow tied to the dust structure.

\begin{figure*}
\centering
\includegraphics[angle=0,width=.49\textwidth]{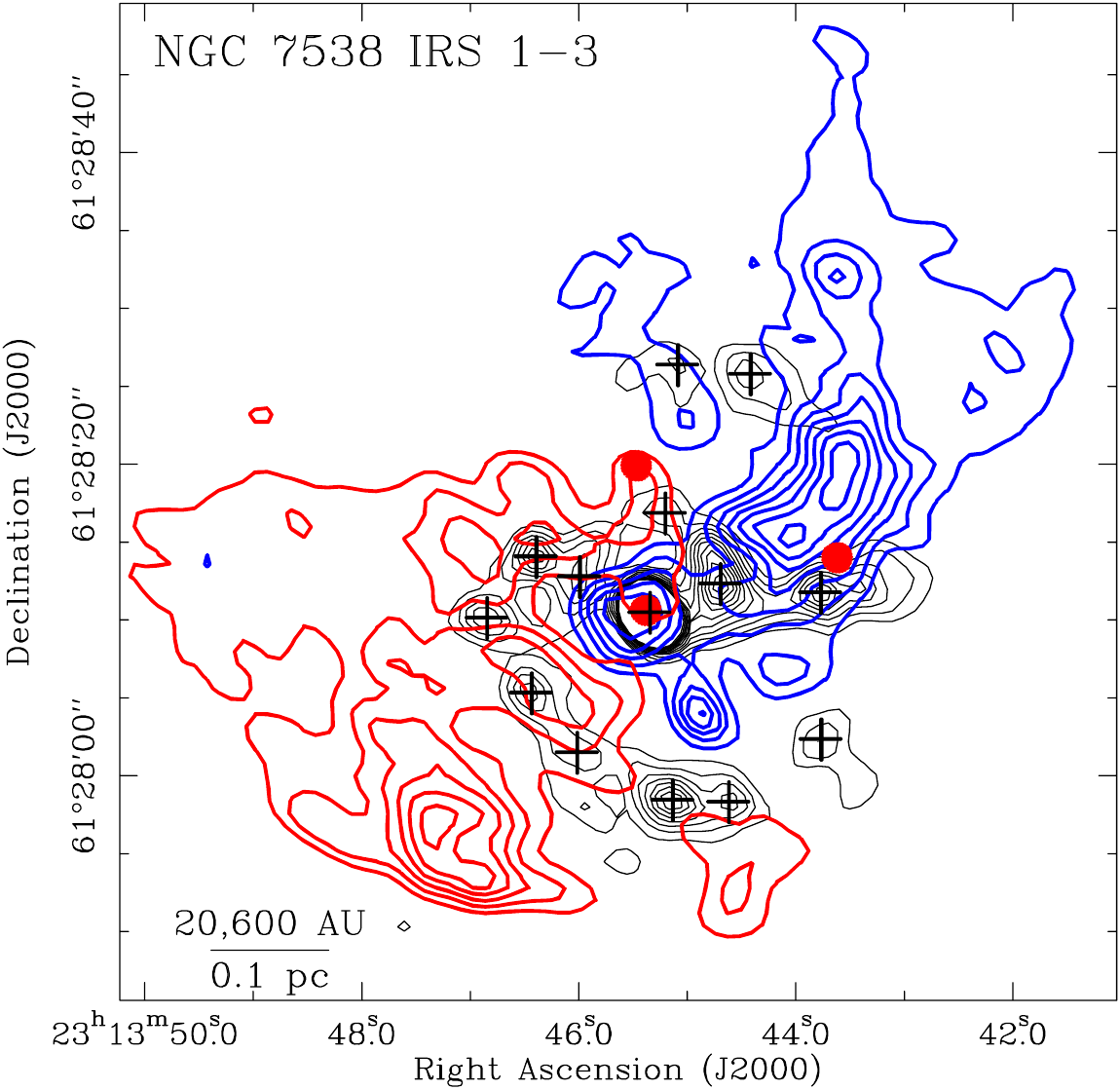} 
\hfill
\includegraphics[angle=0,width=.45\textwidth]{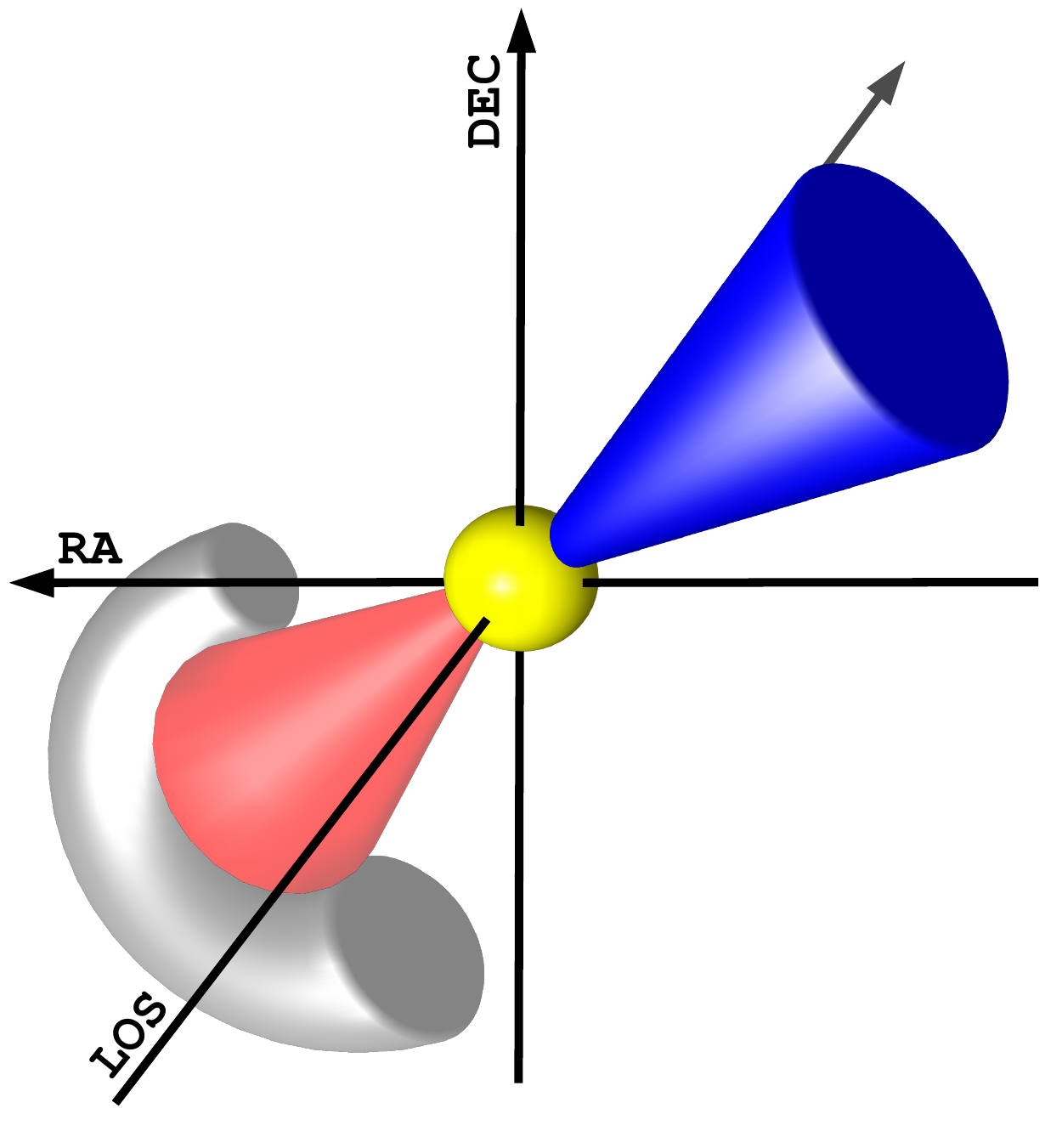} 

\caption{{\it Left panel}: overlayed contours for dust continuum (gray, this work), and blue-- and red--shifted $^{13}$CO~2--1 outflow (blue and red, QZM11). Crosses mark the positions for dusty cores and red dots for IR sources. {\it Right panel}: schematic 3D cartoon of the proposed scenario for the \src--3 complex (Section~\ref{ssec-feedback}). Yellow sphere represents MM1, blue and red cones represent the blue and red outflow lobes, and the gray structure represents the spiral arm. Faded colors represent the structures behind MM1 in the LOS direction.}

\label{fig-dust-outflow}
\end{figure*}

\subsection{Energetics of the spiral arm}\label{ssec-dynamics}

The analysis of the kinematics in Section~\ref{ssec-kinetic} suggests that the spiral arm around \src\ is expanding, although we cannot discard a certain contribution from rotation. This result was already suggested based on CS observations by \citet{kawabe92}.

For the expansion to happen the filament must be gravitationally unbound to the total mass around IRS~1. The combined mass of MM1, MM2, MM4, plus the star embedded in MM1, is 110~\msun. The distance of the filament with respect to IRS~1 is $\sim$13\arcsec\ (0.17~pc at 2.65~kpc). For these quantities, the virialized rotation velocity of the filament is 1.7~\kms, and the escape velocity is 2.4~\kms. The velocity difference of the filament with respect to the central region is in the 1.5--4.5~\kms\ range when projected in the POS, or 9~\kms\ according to the best fitting kinematic model. Therefore, the spiral arm appears to be gravitationally unbound with respect to the massive MM1, MM2, and MM4 cores. We note that the measured mass is a lower limit due to the filtering effect of the SMA.  However, it is required a mass of \tenpow{1.6}{3}~\msun\ within the central $\sim$10\arcsec\ to gravitationally bound the gas moving at 9~\kms.  The single-dish measured mass of the entire 1\arcmin\ clump is \tenpow{3.7}{3}~\msun\ \citep[corrected for the different  $\kappa_{\rm dust}$ and distance used]{Momose01}. It seems unlikely that 40\% of the total mass is accumulated whitin the central 10\arcsec, and hence, it is unlikely that the spiral--arm is gravitationally bound.

Focusing on the filament, the total mass combining MM3, MM6, MM7, MM7b, MM10, and MM12 is 45.8~\msun. Hence, the total gravitational energy of the filament using a radius of 13\arcsec\ is $E_{\rm grav}$=\tenpow{2.6}{45}~erg. In addition, we can derive relevant dynamical parameters from the kinematic model. We assumed for the calculations that the expansion velocity is constant. Then, we considered two different scenarios: ({\it i}) a conservative one where the expansion velocity is the maximum velocity projected in the POS, 4.5~kms, and ({\it ii}) that drawn by the kinematic model that takes into account projection effects and results in a faster velocity, 9~\kms. The conservative scenario delivers an age of $t_{\rm dyn}$$\simeq$\tenpow{3.6}{4}~yr and kinetic energy of $E_{\rm kin}$$\simeq$\tenpow{9.2}{45}~erg, while the deprojected scenario delivers $t_{\rm dyn}$$\simeq$\tenpow{1.8}{4}~yr and $E_{\rm kin}$$\simeq$\tenpow{3.7}{46}~erg. The gravitational well is a factor of 3.5--14 smaller, thus confirming the plausibility of an expanding filament. 

To complete the picture, a crude estimate of the magnetic field energy can be done by multiplying the volume of mass permeated by the field times the overall magnetic pressure ($P_{\rm B}=B_{0}^{2}/(8\pi)$). An approximate area of 150\arcsec$^{2}$ and depth of 5\arcsec\ delivers a magnetic energy of $E_{\rm mag}$=\tenpow{1.3}{45}~erg, negligible for the overall filament dynamics when compared to kinetic energy.

The correlation of the field morphology to the dust morphology can now be tentatively explained: the 3--9 times more energetic expansion motions seem to push away both matter and magnetic field, shaping them in a similar morphology as shown in Fig.~\ref{fig-b-dust-ang}.

\subsection{Formation of the spiral arm through stellar feedback\label{ssec-feedback}}

The large scale expansion motions proposed request a powerful driving source. A plausible origin is the IRS~1 feedback, and in particular from the powerful NW-SW molecular outflow powered by IRS~1 (QZM11). We show in Fig.~\ref{fig-dust-outflow} a comparison of the $^{13}$CO outflow from MM1 to our continuum map. Projected on the POS, it seems that the outflow is perpendicular to the spiral arm, suggesting that it  could be formed by swept material.

Quantitatively, QZM11 derive $E_{\rm outflow} = $\tenpow{4.9}{46}~erg, which is a factor of 1.2--3.7 larger than the combined kinetic, magnetic, and gravitational energy of the filament. In addition, energy losses are expected in the form of turbulence in this complex scenario. The average non-thermal velocity dispersion is $\Delta v^{\rm NT}$=1.8~\kms, which implies an extra \tenpow{1.5}{45}~erg that the outflow can provide. Therefore, it is feasible that the molecular outflow is the energy source responsible for the expansion of the spiral arm.  Furthermore, the outflow momentum for one lobe, \tenpow{2.3}{2}~\msun~\kms, is comparable to the momentum of the spiral arm, $\simeq$\tenpow{2}{2}~\msun~\kms, suggesting that the outflow is the unique cause to set the filament into motion. Finally, the outflow dynamical timescale of \tenpow{2}{4}~yr falls within the filament age range if expanding, suggesting that the two structures have contemporary births.

We can take into account the inclination effects for a better determination of the outflow effect. Based on our kinematic model, we found that the plane containing the logarithmic spiral lies almost parallel to the POS, tilted by $\sim$20\deg\ (70\deg with respect to the LOS, see Table~\ref{tab-toymodel}). This information may help to understand the three-dimensional orientation of the system. One possibility is that outflow and spiral arm flow are parallel, and thus, the gas forming the spiral arm would be pushed by the outflow end. The opposite possibility is that the spiral arm is perpendicular to the outflow, implying that the material would be blown away from the cavity \citep[$\sim$80\deg\ wide: ][]{Kraus06} and driven to the equatorial plane of the system. The aligned scenario implies an outflow lying almost on the POS, with an increase on de-projected outflow velocity by a factor of $\sim$3 and on $E_{\rm outflow}$ by a factor of $\sim$9, while the perpendicular scenario renders an outflow inclined by 20\deg\ with respect to the LOS, and a mild increase of $\sim$6\% in outflow velocity and $\sim$13\% in $E_{\rm outflow}$. A possibly precessing outflow axis may increase the uncertainty in our analysis \citep{Kraus06}. Also, we might consider configurations between the parallel and perpendicular scenarios. The perpendicular case would imply that both structures have opposite velocity directions when projected on the POS. Therefore, the fact that the spiral arm is mostly red-shifted towards the same direction as the large scale outflow seems to favor the aligned scenario. There are other cases in the literature of sweeping up the ambient material as a snow-plow and accumulating it into a shell \citep{anglada95,girart05}.

Consequently, based on the morphology considerations in this section, and on the energy considerations in Section~\ref{ssec-dynamics}, we speculate that the dusty spiral arm is created by the accumulation of matter due to the IRS~1 outflow feedback.

\subsection{Gravitational collapse of the individual cores: a cluster in the making\label{ssec-collapse}}

Three active, bright IR sources indicate that star formation is ongoing in the NGC~7538~IRS~1--3 cluster. Moreover, through high-resolution IR interferometry, \citet{Kraus06} find 18 new faint stars and a NW-oriented, fan-shaped outflow arising from MM1. Interestingly, the positions of the stars are correlated well with the outflow, which they propose is precessing and triggering star formation. In this environment of high interaction, we target the cold dust  to study the possible evolution of the mass reservoir in the cluster.

 The three cores in the central dusty structure seem to be gravitationally dominated against the support forces (Section~\ref{ssec-energy-sources}, note that MM4 is only at the limit). The MM1 core has already formed a still accreting protostar, which has gathered about two thirds of the total mass locally available (star plus dust core system). If the same star-to-core mass ratio applies, MM2 and MM4 will form massive stars of $\sim$24~\msun\ and $\sim$17~\msun, respectively.

\begin{figure}
\includegraphics[width=\columnwidth]{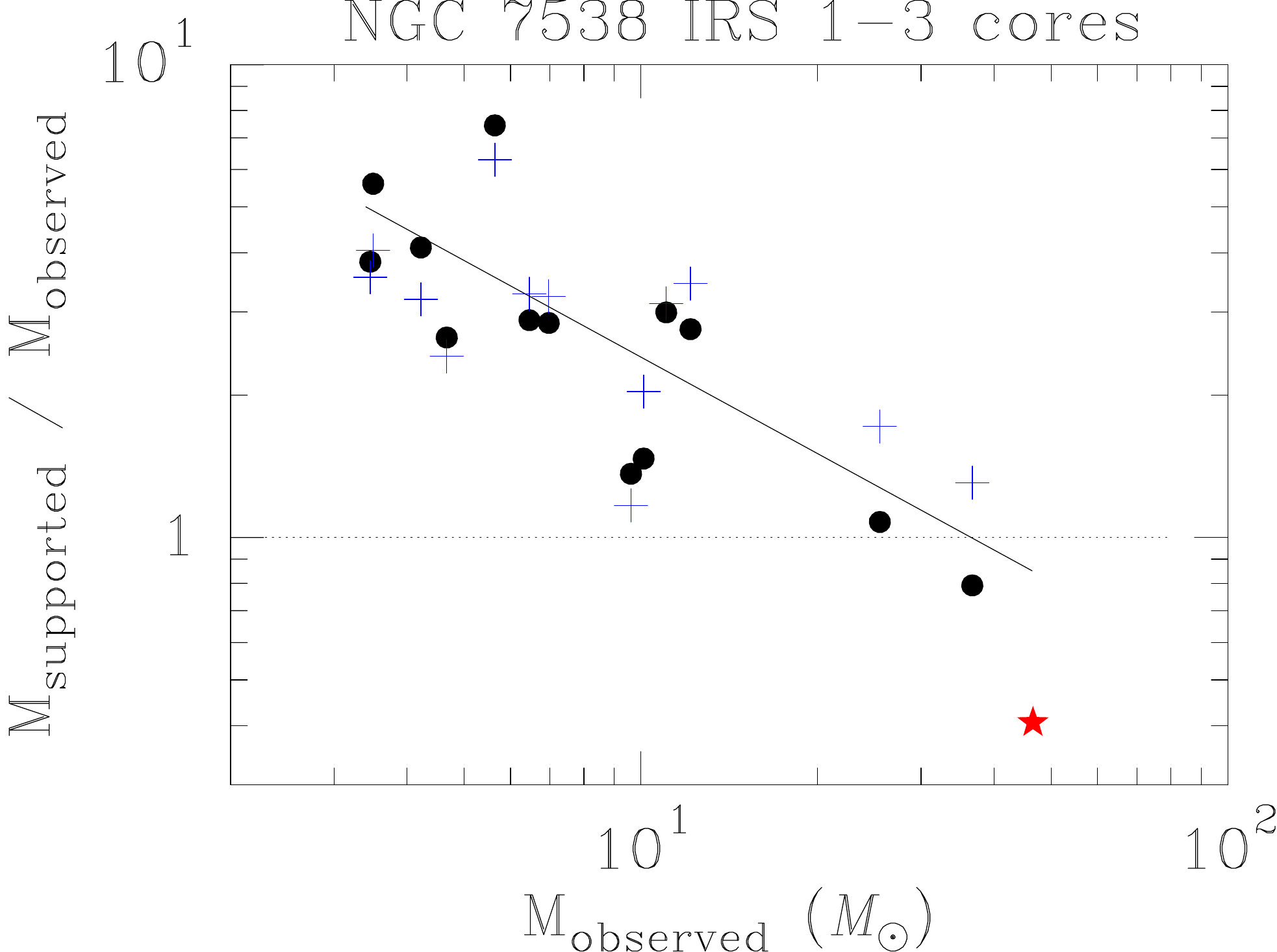}
\caption{Ratio of supported mass to observed mass as a function of the observed mass for the \src--3 dense cores. MM1 is represented by a red star and the other cores by black circles. The solid line shows the best fit to the starless cores while blue crosses show the expected values for critical, magnetized clumps (see Section~\ref{ssec-collapse}). The dotted line shows the limiting value for effective support against gravitational collapse.}\label{fig-msup-mobs}
\end{figure}

The situation is less clear for the dust cores located either in the spiral arm or isolated. The ``mass balance'' analysis shows that the total mass is insufficient to gravitationally bound the cores. Figure~\ref{fig-msup-mobs} shows an extension of the traditional virial parameter analysis with the inclusion of the magnetically supported mass. This parameter is usually fitted by a function of the form $M_{\rm obs}^{a}$ that delivers a typical $a=-0.68\pm0.06$, in agreement with previously reported trends \citep[see e.g.][and derived works]{bertoldi92}. The data can also be fitted by a function of the form $b\left[M_{\rm vir}/M_{\rm obs}\right]^{a}$, where the proportionality constant $b$ carries a physical meaning, varying from 2.06 for self-graviting clumps to 2.9 for pressure-confined clumps \citep{bertoldi92}. The best fit to the cores delivers $b=2.26\pm0.12$, in good agreement with the prediction of $b=2.12$ for magnetized critical cores. To examine the range of applicability of the model, we show as blue crosses in Fig.~\ref{fig-msup-mobs} the expected results using the masses of the clumps, the theoretical $b=2.12$, and the previously derived $a$. The prediction for magnetized critical cores is in good agreement with the magnetically dominated low-mass cores (see also Section~\ref{ssec-energy-sources} and Fig.~\ref{fig-massBalance}). On the contrary, the model prediction is less precise for the more massive cores, where the magnetic field is less important and are less likely to be magnetically critical.

The analyses performed assume cores in isolation and ignore the effects of the highly dynamical environment. In other words, the expansion powered by the outflow may help to pile up material as the filament is expanding through the ISM, and more importantly, it is creating a high external pressure along the filament. This pressure can help gravitation to overcome magnetic and turbulent energy. We can estimate if an external perturbation can have a significant impact on core evolution by using the typical crossing-time $t_{\rm cross}$=$R_{\rm core}/c_{s}$. In average, the filament cores have $R_{\rm core}\simeq$\tenpow{3.6}{3}~AU and $c_{s}\simeq$0.53~\kms, resulting in a typical $t_{\rm cross}$$\simeq$\tenpow{3}{4}~yr. This value is comparable to the $t_{\rm dyn}$ estimates for the filament and outflow (Section~\ref{ssec-dynamics}). Therefore, a perturbation constantly acting for $t_{\rm dyn}$, such as stellar winds, can influence the evolution of the cores in the filament. Since all measures suggest that the outflow is behind the formation of the entire filament (Section~\ref{ssec-feedback}), comparable timescales at a core level make reasonable that the external pressure will also influence the cores.

We speculate that the external pressure from the winds acting for $t_{\rm dyn}$ could trigger the collapse of the cores in the filament, leading to the formation of a group of low-mass stars. This triggered star formation SE of MM1 is supported by the mirrored star formation towards NW, in a more evolved stage of evolution. These stars are older than the MM1 outflow and could have been formed through feedback from the older, more evolved IRS~2 star. This star is associated to a well studied \hd\ region and powers a ``stellar wind bow shock'' \citep{bloomer98}. Consequently, triggered star formation in the \src--3 complex could be an episodic process following the evolution of the most massive stars. Such scenario would generate a small cluster with two stellar groups: ({\it i}) a few central high-mass stars, surrounded by ({\it ii}) a wealth of low-mass stars formed through feedback from the former group.

\section{Conclusions\label{sec-concl}}

We have carried out a molecular, dust, and polarimetric study of the NGC~7538~IRS~1--3 star-forming cluster. We used SMA high angular resolution observations at 880~$\mu$m with the compact configuration. Here, we summarize the main results.

\begin{enumerate}

\item We detect up to 14 dust cores in continuum emission, six of them newly discovered, spanning one order of magnitude in mass (from 3.5~\msun\ to 37~\msun). The brightest core is MM1, associated with IRS~1. IRS~2 and IRS~3 show no continuum counterpart. The dust cores are connected by diffuse gas, and are arranged in two larger scale structures: a central bar containing MM1, MM2, and MM4; and a filamentary spiral arm containing at least 6 cores. The total dust mass is $\simeq$160~\msun, almost equally split between the two large scale structures.

\item We detect \cdo~3--2 and \htcop~4--3 large scale emission sharply tracing the two main large scale structures, unveiling a velocity gradient along the spiral arm. We developed a code to generate synthetic velocity cubes, \texttt{RATPACKS}, and reproduced the velocity gradient through a model of a spiral expanding at 9~\kms\ with respect to the central MM1.

\item We broadly detect polarized emission in the compact cores and in the diffuse extended structures. Based on a statistical analysis, we derive a magnetic field strength of $\simeq$2.5~mG. The orientation of the magnetic field segments is significantly homogeneous along the spiral arm, and it is correlated at an 80\% confidence level to the direction of the dust main axis. This suggests that dust and magnetic field are tightly connected.

\item The spiral arm is gravitationally unbound with respect to the central bar. The gravitational and magnetic field energies combined are a factor of 2.3--9.5 smaller than the kinetic energy. Therefore, it is likely that the dominant expansion is shaping dust and magnetic field into a similar morphology. 

\item The total energy, linear momentum, and dynamic age ($\simeq$\tenpow{4.2}{46}~erg, $\simeq$\tenpow{4}{2}~\msun~\kms, and $\simeq$\tenpow{1.8}{4}~yr) of the spiral arm are compatible with the values of the MM1 outflow by QZM11 when de-projected. Both spiral arm and outflow are red-shifted, hence likely to flow in parallel. Consequently, it seems plausible that the dominant kinetic energy of the spiral arm has its origin in the MM1 outflow, which may be causing its formation in a snow-plow fashion in agreement to our expansion model.

\item We developed the ``mass balance'' analysis that compares collapse vs. support forces, accounting for all the available information on the energetics at core scales. On the one hand, the cores in the central bar seem to be gravitationally unstable, and prone to form massive stars. On the other hand, the combined support forces seem to dominate the cores located in the spiral arm or isolated, with non-thermal motions and magnetic fields being the main agents of support, respectively. However, the dynamically important external pressure from the outflow could trigger the gravitational collapse, and lead to the formation of low-mass stars as reported towards NW to MM1 \citep{Kraus06}.

\item We speculate that the NGC~7538~IRS~1 region is forming a small cluster with a few central high-mass stars, surrounded by a number of low-mass stars formed through proto-stellar feedback.

\end{enumerate}


\begin{acknowledgements}

We thank all members of the SMA staff that made these observations possible. This research has made use of NASA's Astrophysics Data System Bibliographic Services (\texttt{http://adsabs.harvard.edu/}), the SIMBAD database, operated at CDS, Strasbourg, France (\texttt{http://simbad.u-strasbg.fr/simbad/}), and the Splatalogue database for astronomical spectroscopy (\texttt{http://www.splatalogue.net}). We thank the anonymous referee for the useful comments. PF is supported by the Spanish CONSOLIDER project CSD2009-00038. PF and JMG are supported by the Spanish MINECO AYA2011-30228-C03-02, and Catalan AGAUR 2009SGR1172 grants. 

\end{acknowledgements}


\end{document}